\documentclass[aps,pre,twocolumn,groupedaddress,showpacs,preprintnumbers,amsmath,amssymb]{revtex4}

\usepackage{graphics,graphicx} 

\newcommand{\meanval}[1]{\left\langle #1 \right\rangle} 
\newcommand{\InsertFig}[1]{\includegraphics{#1.eps}} 
\newcommand{\setZ}{{\mathord{\mathbb Z}}}
\newcommand{\zu}{\rightarrow}
\newcommand{\dist}{\operatorname{dist}}
\newcommand{\err}{\operatorname{err}} 
\begin{document}

\title{Fast Flat-Histogram Method for Generalized Spin Models}

\author{S. Reynal}
\email{reynal@ensea.fr}
\homepage{http://www-reynal.ensea.fr}
\altaffiliation{Permanent address: E.N.S.E.A., 6 avenue du Ponceau, 95014 Cergy Cedex, France.}
\author{H.T.Diep}
\affiliation{Laboratoire de Physique Th\'eorique et Mod\'elisation, CNRS-Universit\'e de Cergy-Pontoise (UMR 8089), 2 avenue A. Chauvin, F-95302 Cergy-Pontoise Cedex, France}

\date{\today}

\begin{abstract}
We present a Monte Carlo method that efficiently computes the density of states for spin models having any number of interaction per spin.
By combining a random-walk in the energy space with collective updates controlled by the microcanonical temperature, our method yields dynamic exponents close to their ideal random-walk values, reduced equilibrium times, and very low statistical error in the density of states. The method can host any density of states estimation scheme, including the Wang-Landau algorithm and the transition matrix method. Our approach proves remarkably powerful in the numerical study of models governed by long-range interactions, where it is shown to reduce the algorithm complexity to that of a short-range model with the same number of spins.  
We apply the method to the $q$-state Potts chains $(3\leq q \leq 12)$ with power-law decaying interactions in their first-order regime; we find that conventional local-update algorithms are outperformed already  for sizes above a few hundred spins. By considering chains containing up to $2^{16}$ spins, which we simulated in fairly reasonable time, we obtain estimates of transition temperatures correct to five-figure accuracy. Finally, we propose several efficient schemes aimed at estimating the microcanonical temperature. 
\end{abstract}


\maketitle

\section{Introduction}\label{sec:introduction}

Long-range spin models have drawn increasing interest in the last decade, both in the microscopic modeling of a variety of systems ranging from model alloys \cite{GiacominLebowitz1996} to spin glasses \cite{Ford1982} to neural networks \cite{Amit1989}, and as powerful laboratory frame to investigate fundamental issues in the physics of critical phenomena. These include, e.g., the effect of dimensionality \cite{Luijten1999}, the crossover from short-range to long-range behavior \cite{Theumann1985,LuijtenBlote2002,ReynalDiep2004a}, mean-field driven phase transitions \cite{Biskup2005}, and possible connections with Tsallis's non-extensive thermodynamics \cite{Cannas1996, VollmayrLee2001,CannasLapilliStariolo2004}.
Monte Carlo (MC) methods have now gained a prominent role in the investigation of phase transitions in these models \cite{Luijten1995,GlumacUzelac1997_1998b,BayongDiepDotsenko1999,Binder2000,Binder2000a,BinderLuijten2001,UzelacGlumacAnicic2001}. 
In particular, a major breakthrough was recently initiated by the introduction of a (canonical) cluster algorithm able to overcome the algorithm complexity 
inherent to long-range (LR) models, namely, the need to take a huge number of interactions into account at each Monte Carlo step (MCS) \cite{Luijten1995}. In a recent article, we proposed a generalization of this algorithm to simulations in the multicanonical ensemble \cite{Reynal2005a}. It is the goal of the present work to introduce a general and versatile method aimed at embedding any cluster update scheme in a flat histogram algorithm, with special emphasis given to LR spin models.

Whether short- or long-range interactions are considered, canonical MC simulations of long-range spin models suffer indeed from severe shortcomings, the use of cluster updates notwithstanding. First and foremost, models exhibiting first order phase transitions or complicated energy landscapes experience supercritical slowing down \cite{Berg1992}: the time needed for the dynamics to tunnel through free energy barriers grows exponentially with the lattice size, leading to quasi ergodicity breaking and unreliable statistics. Second, the computation of free energies and related thermodynamic quantities is highly involved, and a precise determination of the order of the transition is often intractable. In practice, these shortcomings preclude the use of canonical MC algorithms at first-order transitions except at modest lattice sizes and in the case of weakly first-order transitions.

An efficient approach aimed at overcoming this limitation is the simulation in generalized ensembles \cite{Iba2001,Berg2002}, in particular its multicanonical flavor initially proposed by Berg \cite{BergNeuhaus1991,BergNeuhaus1992,Berg1992}, reconsidered in the context of transition matrix dynamics \cite{Wang1999,SmithBruce1995} and recently revisited by Wang and Landau \cite{WangLandau2001a,WangLandau2001b}. The key-idea here is to artificially enhance rare events corresponding to local maxima in the free energy, by feeding the Markov chain with an appropriate distribution $w(E)$. In the {\em multicanonical ensemble}, $w(E)$ is set to the inverse of the density of states, so that the resulting dynamics is a random walk in the energy space that yields a flat histogram of the energy. Other ensembles have been proposed in the last decade, including the $1/k$ ensemble, which enhances low-energy states \cite{HesselboStinchcombe1995}, and very recently, the optimal ensemble, which aims at optimizing the distribution $w(E)$ with respect to the local diffusivity of the random walker, so that tunneling times are minimized \cite{Trebst2004,WuTrebst2004}. While still broad, histograms engendered by these last ensembles are no longer flat; in the optimal ensemble for instance, the histogram is slightly peaked around the critical region, so that the larger time spent by the random walker inside the critical region compensates the lower diffusivity in this region.

When implemented through local (i.e., single-spin) updates \cite{ReynalDiep2004a}, simulations in the multicanonical ensemble suffer, however, from two serious hurdles. First, while tunneling times --- measured in Monte Carlo steps (MCS) --- are reduced from an exponential to a power law $\tau \sim L^z$ of the lattice size, 
the dynamic exponents $z$ are still substantially higher than the ideal value $z \sim D$ that should be expected from the dynamics of a random walker \cite{JankeKappler1995b,DayalTrebst2004}. This observation, as we will witness in this article, applies equally well to effective autocorrelation times and to equilibrium times; this represents a serious hindrance in terms of scalability, in particular whenever a higher precision is desired and large amounts of decorrelated data need to be gathered. In this respect, it is important to mention that correlations between successive measurements do not only have an impact on the statistical efficiency of multicanonical production runs, yet also represent a source of systematic error regarding the estimation of the density of states \cite{Zhou2003}. 
A second impediment to the scalability of local-update implementations specifically relates to long-range models. Here, the very presence of long-range interactions makes the computation of the energy --- an essential ingredient of multicanonical methods --- a very time consuming operation, namely, one associated with an $O(L^{2D})$ algorithm complexity. As a result, the demand on CPU time needed to generate perfectly decorrelated statistics grows as $L^{z+2D}$, with $z > D$.

In this article, we present a Monte Carlo method which successfully tackles these issues by performing simulations in the multicanonical ensemble using collective updates. 
Our methods combines the fast-decorrelating capabilities of cluster algorithms with the versatility of flat-histogram methods in an efficient and straightforward way, and with wide applicability in view. In particular, it can be readily combined with any iteration scheme dedicated to the estimation of the density of states, e.g., Wang-Landau's method \cite{WangLandau2001a} or transition matrix algorithms \cite{Wang1999}. Additionally, while our method is presented here in the context of long-range spin models, where it gives drastic improvements over commonly used methods, it is perfectly applicable to any class of models for which a cluster algorithm exists in the canonical ensemble. 

Noteworthy enough, embedding a collective update scheme in a multicanonical algorithm is not straightforward, however, due to the fundamentally {\em non-local} nature of the multicanonical weight $w(E)$. Indeed, cluster algorithms depend heavily upon particular symmetries of the model Hamiltonian, which $w(E)$ does not keep track of; in particular, there is no longer a canonical temperature. With simulations of spin models with nearest-neighbors interactions in view, several attempts have been made at combining cluster updates with multicanonical methods in some way or another during the last decade: the multibond algorithm \cite{JankeKappler1995a,JankeKappler1995b,Janke1998,CarrollJankeKappler1998} or variants thereof targeting Wang-Landau's algorithm \cite{Yamaguchi2003, WuTrebst2004} simulate the model in its spin-bond representation; Rummukainen's hybrid-like two-step algorithm lumps together a microcanonical cluster algorithm and a multicanonical daemon refresh \cite{Rummukainen1993}.
As opposed to these, however, our method relies on a cluster-building process which simply depends on the microcanonical temperature of the current configuration --- a quantity that may be readily derived from the estimated  density of states --- in order to determine appropriate bond probabilities. In particular, it does not require prior knowledge of the transition temperature, as is the case in the multibond method. We further show that our approach makes it particularly straightforward to incorporate two optimization schemes dedicated to LR models \cite{Luijten1995,KrechLuijten2000}, which cut down the algorithm complexity from $O(L^{2D})$ to $O(L^D\ln L^D)$. As a result, the total demand on CPU time with respect to uncorrelated data is reduced to approximately $L^{2D} \ln L^D$, since cluster updates also lower $z$ to around $D$; where LR models are concerned, the benefit of cluster updates is thus clearly twofold. Let us also mention that, as a by-product, using cluster updates provides improved estimators for the statistical moments of the order parameter \cite{Uzelac2002} and for spin-spin correlation functions; for instance, the last quantity can be better estimated by counting the fraction of time two given sites belong to the same cluster \cite{Niedermayer1988,Niedermayer1996}. Further interesting information, including information connected with fractal geometry, may also be gleaned from cluster statistics \cite{Asikainen2003,Janke2005}.

Overall, the sharp reduction of the computer load brought about by our method allowed us to study $q$-state Potts chains with $1/r^{1+\sigma}$ interactions containing up to $2^{16}$ spins in a few days on a modern Intel-based computer. It must be noted that, with standard multicanonical methods based on single-spin updates, such huge sizes are simply intractable, since the largest size of $2^{16}$ investigated in this work would demand several months of computation.
As regards dynamic performance, we obtain a substantial reduction in the dynamic exponent, from e.g., $z\sim 1.35(3)$ to $z\sim 1.05(1)$ for $q=6$ and $\sigma=0.7$. We also show that our method produces faster equilibration, lower effective autocorrelation times, and --- where implementations based on the Wang-Landau algorithm are concerned --- lower statistical errors in the estimate of the density of states, e.g., of nearly an order of magnitude for $q=6$, $\sigma=0.9$ and $L=512$ spins.
As a result, we obtain estimates of transition temperatures that have a noticeably higher precision than those obtained using local updates \cite{ReynalDiep2004a} or standard canonical methods \cite{GlumacUzelac1997_1998b,BayongDiepDotsenko1999}.  
Finally, in order to check that our method did not produce systematic errors, we performed several simulations of the two-dimensional seven- and ten-state Potts models with nearest-neighbor interactions and sizes up to $L=256 \times 256$. We obtain dynamical exponents close to the ideal random-walk value $z\sim 2$. Although computed from rather modest statistics, our estimate of the interfacial free energy for the largest size reaches a precision of four digits. 
In this respect, our method compares perfectly well with other methods operating in the multicanonical ensemble, and represents an alternative way for short-range spin models.

The layout of this article is as follows. In Sec.~\ref{sec:algorithm}, after briefly reviewing some  prominent features of multicanonical methods, we explain how we combine a multicanonical weighting with collective updates, with special emphasis given to the detailed balance equation.  Section~\ref{sec:lr_optimization} addresses optimizations dedicated to long-range models. Numerical results regarding the dynamic characteristics of our method are presented in Sec.~\ref{sec:results_dyn_behaviour}. In Sec.~\ref{sec:NN_model}, we compare our results for the two-dimensional Potts model with nearest-neighbor interactions, with exactly known results, and section~\ref{sec:LR_model} is devoted to the investigation of the precision of our method in the context of the long-range Potts chain with power-law decaying interactions. Overall, we pay particular attention to comparison with other algorithms operating in the multicanonical ensemble, especially in terms of tunneling rates, dynamical exponents and estimates of thermodynamical averages. Finally, we discuss several procedures aimed at estimating the microcanonical temperature, and in particular, how we can efficiently combine our method with the transition matrix approach.

\section{A method to embed cluster updates in a flat histogram algorithm}\label{sec:algorithm}
Monte Carlo simulations are based on the generation of a Markov chain of configurations $\{\sigma_i\}$, where each configuration is assigned a weight $w[E(\sigma_i)]$ corresponding to the probability distribution one wishes to sample. 
In canonical simulations, i.e., carried out at a fixed inverse temperature $\beta$, one chooses a Boltzman weight $w[E(\sigma_i)] = e^{-\beta E(\sigma_i)}$, thus thermodynamical averages are straightforwardly obtained by computing the appropriate moments of the data accumulated at the given temperature. On the other hand, reweighting methods based on multihistogramming \cite{FerrenbergSwendsen1989} are hampered at large lattice sizes by the narrowness of the energy window that is sampled, let alone additional supercritical slowing down. 
In the multicanonical ensemble, one allows the dynamics to jump across free energy barriers and, from a more general standpoint, to sample wide energy windows, by producing a flat energy distribution over the energy range of interest for the problem at hand. This is formally carried out by setting $w(E)=e^{-S(E)} \propto 1/n(E)$, where $n(E)$ is the density of states and $S(E)$ is the microcanonical entropy. This in effect leads to $N(E) \propto n(E) w(E) = \mbox{const.}$ for the number of visits to energy $E$. Since the density of states is obviously a priori unknown, $w(E)$ is estimated using an iterative procedure initially fed from, e.g., a canonical guess $w(E)=e^{-\beta_0 E}$ at a carefully chosen inverse temperature $\beta_0$, a flat guess $w(E)=1$, or --- whenever feasible --- a properly scaled estimate obtained at a smaller lattice size. Thermodynamic quantities that depend solely on the energy, like the specific heat or Binder cumulants, can then be computed directly from the estimated density of states. Other quantities, e.g., those depending on the order parameter, are obtained through a reweighting procedure based on data gathered during an additional production run.

Historically, Berg's recursion scheme \cite{Berg1996, Berg2000} was the first iteration procedure specifically dedicated to multicanonical simulations.
It consists in accumulating histogram entries of the energy during each iteration run, and updating $w(E)$ from the histogram of the energy obtained in a previous iteration run, until eventually the histogram becomes flat up to a given tolerance. Entropic sampling \cite{Lee1993} more or less boils down to the same key principle. Both methods suffer, however, from poor scalability.
Looking at this issue from a slightly different angle, the recently proposed Wang-Landau acceleration method \cite{WangLandau2001a,WangLandau2001b} updates $w(E)$ in real-time during the course of the simulation, performing independent random walks in distinct energy ranges. Since modifying the weight of the Markov chain during a simulation breaks detailed balance, the amount by which $w(E)$ is modified during a given iteration is decreased from one iteration to the other until it reaches a negligible value; hence detailed balance is approximately restored in the last step of the iteration scheme. 
In this respect, an original approach aimed at reducing the statistical error in the density of states was recently proposed by Yan and de Pablo \cite{YanDePablo2003}, whereby the density of states is obtained by integrating an instantaneous temperature computed from configurational information or from a so-called multimicrocanonical ensemble.
Finally, a large class of iteration schemes have been proposed that are based on matrices of transition probabilities \cite{SmithBruce1995,Chan1993,Wang1999,Wang2000,Wang2002} or a combination thereof with Wang-Landau's algorithm \cite{Shell2003,Surungan2004}.
 Here, the density of states is computed through a so-called broad histogram equation involving infinite temperature transition matrices, where transition matrices keep track of the microcanonical average of the number of potential moves from one energy levels to another (Sec.~\ref{sec:estimating_betaE} gives more details on how our method can efficiently capitalize on transition matrices). Historically, procedures based on transition matrices were termed \textit{flat histogram methods} in order to distinguish them from Berg's multicanonical method, although both approaches in effect yield a flat, broad histogram. To sum up, the main benefit of multicanonical methods is twofold: first, a wide energy range is sampled, irrespective of the presence of free energy barriers; second, the methods yield a direct estimate of the density of states.

A local-update implementation of a multicanonical algorithm may consist in updating a single spin at a time and accepting the attempted move from state $a$ to state $b$ with a probability given by $W(a\zu b) = \min[1,e^{S(E_a)-S(E_b)}]$. We now show that the microcanonical temperature $\beta(E)$ defined as $dS(E)/dE$ is a relevant quantity for the acceptance rate of this process. Denoting $E_b=E_a+\epsilon$, we expand the probability $W(a\zu b)$ for small $\epsilon$, and obtain $W(a\zu b) \sim \min[1,e^{-\beta(E_a)\epsilon}]$. This shows that, for small enough energy changes, the dynamics is equivalent to that of a canonical simulation at an inverse temperature $\beta(E)$. Our departure point for a collective-update implementation in the multicanonical ensemble is thus to build clusters of spins with the same bond probabilities as would be given by a canonical simulation at inverse temperature $\beta(E)$. 

Although our algorithm may be equally well applied to other spin models, e.g., models incorporating disorder or exhibiting a continuous symmetry, we now consider, for the sake of clarity, a generalized ferromagnetic spin model with a $\setZ_q$ symmetry, whose Hamiltonian reads $H = - \sum_{i < j} J_{ij} \delta_{\sigma_i,\sigma_j}$. Here $J_{ij}>0$ and the $\sigma_i$ variables can take on integer values between $1$ and $q$. 
Taking guidance from Swendsen-Wang's cluster algorithm \cite{SwendsenWang1987}, we start from an empty bond set, consider each pair of spins $\{\sigma_i,\sigma_j\}$ in turn, and activate a bond between them with a bond probability given by $\pi_{ij}(E_a) = \delta_{\sigma_i,\sigma_j} \left[ 1 - e^{-J_{ij} \beta(E_a)} \right]$, where $E_a$ is the current lattice energy and $\beta(E_a)$ the inverse microcanonical temperature at energy $E_a$. Efficient ways of estimating $\beta(E)$ are considered later on in Sec.~\ref{sec:estimating_betaE}. Then, we identify clusters of connected spins using, e.g., a multiple-labeling scheme \cite{HoshenKopelman1974}, draw a new spin value at random for each cluster, and accept the attempted move with an acceptance probability $A_{\mbox{flip}}(a\zu b)$ which ensures that detailed balance is satisfied. The derivation of this probability may be carried out in the following way. First, the total acceptance probability $W(a\zu b)$, i.e., the quantity that enters detailed balance in such a way that $e^{-S(E_a)} W(a\zu b)=e^{-S(E_b)} W(b\zu a)$, is split into two terms $P(a\zu b)$ and $A_{\mbox{flip}}(a\zu b)$ representing a {\em proposed update probability} and an {\em acceptance probability} for the proposed update, respectively. It is straightforward to show that the choice $A_{\mbox{flip}}(a\zu b) = \min \left[1, \frac{P(b\zu a)}{P(a\zu b)} e^{S(E_a)-S(E_b)} \right]$ satisfies the detailed balance equation.
Let us denote $\mathcal{B}$ the set of active bonds over the complete graph $\mathcal{G}$ engendered by all possible interactions: the {\em proposed update probability} is given by the probability to construct a given set $\mathcal{B}$ from an empty bond set, i.e.,
$$
P(a\zu b) = \prod_{b_{ij}\in\mathcal{B}} \pi_{ij}(E_a) \prod_{b_{ij}\in \mathcal{G} \backslash \mathcal{B}} [1-\pi_{ij}(E_a)].
$$
After simplification, we obtain for $\frac{P(b\zu a)}{P(a\zu b)}$,
$$
e^{\beta(E_b) E_b - \beta(E_a) E_a} 
\prod_{b_{ij}\in\mathcal{B}} \frac{e^{J_{ij} \beta(E_b)} - 1}{e^{J_{ij} \beta(E_a)} - 1};
$$
whence
\begin{equation}
A_{\mbox{flip}}(a\zu b)=\min\left[ 1, 
\frac{e^{\alpha(E_a)}}{e^{\alpha(E_b)}}
\prod_{b_{ij}\in\mathcal{B}} \frac{p_{ij}(E_b)}{p_{ij}(E_a)}
\right],
\label{eq:cluster_flip_acceptance_rate}
\end{equation}
where $\alpha(E) = S(E) - \beta(E) E$
and $p_{ij}(E) = e^{J_{ij} \beta(E)} - 1$.
This expression can be further simplified if we consider long-range models whose coupling constant depends only on the distance between spins, i.e., $J_{ij} = J(l)$, where $l=\dist(i,j)$. We have for $A_{\mbox{flip}}(a\zu b)$:
\begin{equation}
	A_{\mbox{flip}}(a\zu b) = \min \left[1, \frac{e^{\alpha(E_a)}}{e^{\alpha(E_b)}} \prod_{l>0} \left[ \frac{p_l(E_b)}{p_l(E_a)}\right]^{B(l)} \right],
\label{eq:cluster_flip_acceptance_rate_lr}
\end{equation}
where $B(l)$ stands for the number of bonds of length $l$.
It is worth mentioning that, if one looks at this equation from the standpoint of canonical simulations at inverse temperature $\beta_0$, we have $w(E)=e^{-\beta_0 E}$; whence $\beta(E)=\beta_0$ and $\alpha(E)$ does no longer depend on $E$. As a result, the acceptance rate $A_{\mbox{flip}}(a\zu b)$ is equal to $1$ and we are back to the original Swendsen-Wang algorithm.

It is also crucial to underline that it is the microcanonical temperature, i.e.,  the lattice energy in the first place, which entirely governs the construction of clusters; indeed, for a given lattice configuration at energy $E$, bonds are placed as if the model were simulated at its microcanonical temperature using a Swendsen-Wang algorithm. As a result, cluster bond probabilities change continuously as the lattice configuration walks along the available energy range of the random walk, so that, e.g., small clusters are built in the upper energy range and conversely large clusters in the lower energy range.

\section{Optimization for long-range models}\label{sec:lr_optimization}

\subsection{Computing the lattice energy through FFT acceleration}

As is apparent in Eq.~(\ref{eq:cluster_flip_acceptance_rate}), determining the acceptance rate of a cluster flip demands that we compute the energy of the new (attempted) lattice configuration, which for long-range models is an $O(L^{2D})$ operation. This is similar to the local-update case, where performing one MC step, i.e., updating $L^D$ spins subsequently, takes a CPU time proportional to the square of the number of spins, seeing that $L^D$ operations are needed after each single spin update to compute the new partial energy between the updated spin and the rest of the lattice. Recently, Krech and Luijten proposed an algorithm that is able to compute the energy of a model governed by translation invariant interactions in $O(L^D\ln L^D)$ operations \cite{KrechLuijten2000}. This method leans on the convolution theorem and the Fast Fourier Transform (FFT), for which numerous efficient radix-based implementations are available. As a result, updating the lattice configuration \textit{globally} rather than a single spin at a time allows us to cut the $O(L^{2D})$ complexity down to an $O(L^D \ln L^D)$ one. A crucial point to be noted here is that this reduction is absolutely intractable with single-spin updates, owing to the very reason that the energy would have to be computed anew after each single-spin update; this requires $L^{D}$ operations, and an FFT algorithm would output no gain at all.

Let us assume that we can write down the model Hamiltonian as a sum of dot products, i.e., $H= -\frac12\sum_{i \neq j} J_{ij} \vec{S}(i)\cdot\vec{S}(j)$, with $J_{ij}$ invariant by translation. This is straightforwardly done when $q=2$, since in this case the dot product reduces to a product of scalar Ising spins. 
As we will witness in a moment, the presence of a delta Kronecker symbol in the Hamiltonian whenever $q>2$ requires, however, a minor transformation of the Hamiltonian. For simplicity, we consider hereafter a one-dimensional lattice with an interaction $J(l)$ depending on the distance $l$ between spins. The line argument is similar in higher dimensions, with the sole exception that multidimensional Fourier transforms are then performed. The Discrete Fourier Transform (DFT) of the spin sequence $\{\vec{S}(l)\}_{l=1\dots L}$ reads
$$
	\tilde{\vec{S}}(k) = \sum_{l=0}^{l=L-1} \vec{S}(l) e^{-i 2\pi k l/L},
$$
and reciprocally,
$$
	\vec{S}(l) = \frac{1}{L} \sum_{k=0}^{k=L-1} \tilde{\vec{S}}(k) e^{i 2\pi k l/L}.
$$
Similarly, we define the DFT of the sequence of coupling constants $\{J(l)\}$ as 
$$
	\tilde{J}(k) = \sum_{l=0}^{l=L-1} J_{pbc}(l) e^{-i 2\pi k l/L},
$$
where $J_{pbc}(l)$ incorporates the effect of Infinite Image Periodic Boundary Conditions \cite{CannasLapilliStariolo2004}, that is, $J_{pbc}(l) = \sum_{m=-\infty}^{+\infty} J(l+mL)$; for algebraically decaying interactions, this sum can be exactly expressed in terms of Hurwitz functions \cite{ReynalDiep2004a}.
We diagonalize the original Hamiltonian $H$ by rewriting it in terms of the $\tilde{J}(k)$ and $\tilde{\vec{S}}(k)$,
$$
	H = -\frac{1}{2L} \sum_{k=0}^{k=L-1} \tilde{J}(k) \tilde{\vec{S}}(k) \cdot \tilde{\vec{S}}(-k),
$$
where it should be emphasized that $\tilde{\vec{S}}(-k)$ and $\tilde{\vec{S}}(k)$ are complex conjugates, since the original vectors $\vec{S}(l)$ have real coordinates. By relying on an FFT radix-2 algorithm, the task of computing the lattice energy is consequently reduced to $O(L \ln L)$ operations.

For $q>2$, the Kronecker delta symbol in the Hamiltonian unfortunately rules out the previous diagonalization. 
One way to resolve this issue is to map the $q$-state Potts model onto a $(q-1)-$dimensional vector model, so that the Kronecker delta function in the original Hamiltonian is turned into a dot product. We define a one-to-one mapping between each Potts spin $\sigma=1\dots q$ and a unit-length vector $\vec{S}^{(\sigma)}$ belonging to a $(q-1)$-dimensional hypersphere, so that 
$\vec{S}^{(\sigma)} \cdot \vec{S}^{(\sigma')} = \frac{q \delta_{\sigma,\sigma'}-1}{q-1}$. It is straightforward to prove that $\sum_\sigma \vec{S}^{(\sigma)}=0$, and that 
$$
	H = \frac{q-1}{2q} \sum_{i \neq j} J(i-j) \vec{S}^{(\sigma_i)}\cdot\vec{S}^{(\sigma_j)}
	+ \frac{1}{q} \sum_{i<j} J(i-j).
$$
In the case of the three-state model, this transformation is equivalent to mapping Potts variables onto the complex plane, i.e., $\sigma \zu S^{(\sigma)} = e^{i 2\pi (\sigma-1) /3}$, and writing the dot product 
$\vec{S}^{(\sigma_i)}\cdot\vec{S}^{(\sigma_j)}$ as $Re\{S^{(\sigma_i)}S^{(\sigma_j)*}\}$. In this case, the term $\tilde{\vec{S}}(k) \cdot \tilde{\vec{S}}(-k)$ becomes $|S(k)|^2$, where $S(k)$ is the DFT of the sequence of (complex) variables $\{S^{(\sigma)}\}$. This reduces by one the number of $O(L)$ operations required, since computing a dot product is no longer required.

For $q>3$, spin vectors on the $(q-1)-$dimensional hypersphere may be determined by using hyperspherical coordinates in $D=q-1$ dimensions, i.e., for the $i$th vector $\vec{S}^{(i)}$ (with $1 \leq i \leq q$),
\begin{align*}
	x_1^{(i)} &= \sin \theta_1^{(i)} \sin \theta_2^{(i)} \ldots \sin \theta_{q-3}^{(i)} \sin \theta_{q-2}^{(i)} \\
	x_2^{(i)} &= \sin \theta_1^{(i)} \sin \theta_2^{(i)} \ldots \sin \theta_{q-3}^{(i)} \cos \theta_{q-2}^{(i)} \\
	x_3^{(i)} &= \sin \theta_1^{(i)} \sin \theta_2^{(i)} \ldots \cos \theta_{q-3}^{(i)} \\
	& \ldots \\
	x_{q-2}^{(i)} &= \sin \theta_1^{(i)} \cos\theta_2^{(i)} \\
	x_{q-1}^{(i)} &= \cos\theta_1^{(i)} 
\end{align*}
We initially set $\theta_i^{(i)}=0$ for $1\leq i \leq q-2$, $\theta_j^{(i)}=\alpha_j$ for $j<i\leq q$ and $1\leq j \leq q-3$, and $\theta_{q-2}^{(q-1)}=-\theta_{q-2}^{(q)}=\alpha_{q-2}$. There remains $q-2$ angles $\alpha_j$ to be determined from $q-2$ equations $\vec{S}^{(i)}\cdot\vec{S}^{(i+1)}=-1/(q-1)$ with $1 \leq i \leq q-2$, from where we obtain $\alpha_1=\arccos \frac{-1}{q-1}$, $\cos\alpha_{j+1}=\frac{\cos\alpha_j}{1+\cos\alpha_j}$, and thus by induction $\cos\alpha_j=\frac{-1}{q-j}$. After a bit of algebra, we find 
$\vec{S}^{(1)} = (0, \ldots, 0,1)$, and
\begin{align*}
	\vec{S}^{(i)} = (\underbrace{0, \ldots, 0}_{q-1-i\mbox{ terms }}, & \sqrt{\frac{q(q-i)}{(q-1)(q-i+1)}}, \\
	&\{ x^{(i)}_{q-1-i+j} \} _{1<j<i} , \frac{-1}{q-1}) \\
\end{align*}
for $1<i<q$, where the $(q-1-i+j)$th coordinate reads
$$
	x^{(i)}_{q-1-i+j} = -\sqrt{\frac{q}{(q-1)(q-1-i+j)(q-i+j)}}.
$$
$\vec{S}^{(q)}$ and $\vec{S}^{(q-1)}$ differ only in the sign of their first coordinate $x_1$.
Once these vectors have been computed for a given $q$, which may be done on start-up, determining the lattice energy  requires, first computing the DFT $\tilde{S}_j(k)$ of each sequence of coordinates $\{\vec{S}(l)\cdot\vec{S}^{(j)}\}_{l=1,\ldots,L}$, and then evaluating the double sum $\sum_{k=0}^{k=L-1} \sum_{j=1}^{q} \tilde{J}(k) |\tilde{S}_j(k)|^2$. As a result, the whole operation is associated with a $O(q L\ln L)$ complexity --- or in general $O(q L^D\ln L^D)$ ---, provided the implementation relies on a FFT radix algorithm. As a by-product, it should be noted that once the Fourier components have been computed, it is straightforward to derive the Fourier transform of the spin-spin correlation functions at any inverse temperature $\beta$ from $\tilde{g}_\beta(k)=1/L \meanval{\sum_{j=1}^q | \tilde{S}_j(k)|^2}_\beta$, where the mean value is obtained from a reweighting procedure. At large lattice sizes, the requirement that $L$ Fourier components be stored at each MCS may constitute a significant challenge in terms of computer memory; in this case, a practical work-around consists in computing microcanonical averages for each energy level visited during the simulation, and then to perform the reweighting procedure directly from these microcanonical averages. In the case of long-range interactions, careful attention must be paid, however, to the influence of the discretization of the energy axis in terms of systematic error. 

\subsection{Efficient cluster construction for long-range interactions decaying with the distance}

For long-range spin models, building a new cluster at each MCS takes of order $L^{2D}$ operations, since $L^D(L^D-1)/2$ pairs of spin are considered in turn for bond activation.
When interactions decay with distance, the probability of adding a bond between two spins falls off quite rapidly as the distance between them increases. A significant amount of time during the construction of the cluster is thus wasted because an overwhelming number of bonds are considered for activation which have only a negligible probability to be activated. Even in the case of interactions decaying as $1/|i-j|^{1+\sigma}$ with $\sigma$ close to $0$, does the bond count never exceed a few percent of the whole number of available bonds. In this respect, switching from a local- to a global- update scheme might well be an ill-fated choice as the gain in terms of autocorrelation time is spoiled by the exceedingly time consuming construction of the cluster. However, an efficient construction method was proposed by Luijten and Bl\"ote in the recent past \cite{Luijten1995}, with an efficiency that is independent of the number of interactions per spin, and a CPU demand that scales roughly as $L^D$. The rationale behind this method is to use cumulative probabilities, whereby instead of considering each spin in turn for addition to a given cluster, it is the index of the next spin to be added which is drawn at random.
We now give a sketchy outline of the method in the context of long-range chains.  Extensive details may otherwise be found in \cite{Luijten1995,Luijten2000}. First of all, the probability to add a bond is split up into two parts, namely, (i) a provisional probability $\pi_l(E)$ (hereafter simply denoted $\pi_l$) depending on the distance $l=|i-j|$ between spins and on the lattice energy $E$, and (ii) a factor $f(\sigma_i,\sigma_j)$ controlled by the spin values, e.g., a Kronecker delta symbol in the case of a Potts model. If $0$ denotes the index of the current spin to which we are adding bonds (i.e., spin indices are considered to be relative to the current spin), then the provisional probability of skipping $k-1$ spins and bonding the current spin with a spin at position $k>0$ is given by 
$P_0(k) = \prod_{m=1}^{k-1} (1-\pi_m) \pi_k$. From there, one builds a table of cumulative probabilities 
$C_0(j_1)=\sum_{k=1}^{j_1} P_0(k)$ for all $j_1>0$, so that the index $j_1$ of the spin to be bound with current spin $0$ is obtained by first drawing a random number $0<r<1$ and then reading out $j_1$ from the table, i.e., $j_1$ is such that $C_0(j_1-1) < r < C_0(j_1)$. Standard binary-search algorithms may be used for this purpose. Last, a bond is activated between spins $0$ and $j_1$ with a probability $f(\sigma_0, \sigma_{j_1})$, and we proceed further with the computation of the index $j_2>j_1$ of the next spin to be bound with current spin $0$. The corresponding provisional probability thus becomes $P_{j_1}(k) = \prod_{m=j_1+1}^{k-1} (1-\pi_m) \pi_k$, and the cumulative probabilities read
$C_{j_1}(j_2)=\sum_{k=j_1+1}^{j_2} P_{j_1}(k)$. The same procedure is repeated for $\{j_3,j_4,\ldots\}$ until we draw a $j_\alpha>L$, in which case we jump to the next current spin, which in a one-dimensional model is the nearest-neighbor of the previous current spin. In addition, there are two formulas which make it easier to compute cumulative probabilities: first, one can show that $C_0(j) = 1 - \exp[-\beta(E) \sum_{k=1}^{j} J(k)]$, where $E$ is the energy of the current configuration, and second, the cumulative probabilities $C_{j_{\alpha}}(j_{\alpha+1})$ can be straightforwardly derived from the $C_{0}(j)$ coefficients through the relation $C_{j_{\alpha}}(j_{\alpha+1}) = \frac{C_0(j_{\alpha+1})-C_0(j_{\alpha})}{1-C_0(j_{\alpha})}$. It follows from the last relation that, instead of building a look-up table for each $C_{j_{\alpha}}(j_{\alpha+1})$, we may as well draw a random number $0<r<1$, transform it to $r' = r [1 - C_0(j_{\alpha})]+ C_0(j_{\alpha})$, and choose the next spin to be added from the relation $C_0(j_{\alpha+1}-1) < r' < C_0(j_{\alpha+1})$. 
In practice, we thus simply need to compute a single look-up table filled with $\sum_{k=1}^{j} J(k)$ for each $j$ at the beginning of the simulation, from where we will derive the $C_0(j)$ coefficients at each new MCS corresponding to a lattice configuration with a given energy $E$. This last task requires of order $L^D$ operations. To sum up, the construction of each cluster thus consists in choosing a "current" spin amongst $L-1$ possible spins in turn, e.g., starting from the leftmost one, and then activating bonds between the current spin and other spins located to its right by drawing a random number, scaling it, and selecting the bond indices from a look-up table containing the $C_0(j)$ coefficients at energy $E$. Once each spin has been considered as a current spin, a cluster multiple labeling technique can eventually be used to identify every set of spins actually belonging to the same cluster \cite{HoshenKopelman1974}.

\section{Numerical tests of algorithm performance}\label{sec:results_dyn_behaviour}
In this section, we address the performance of our algorithm in terms of dynamical behavior. Since our work focuses mainly on long-range spin models, we decided to perform intensive numerical tests on the one-dimensional $q$-state Potts chain with LR interactions $1/|i-j|^{1+\sigma}$ decaying as a power law of the distance between spins. The rich phase diagram of this model, and the fact that several numerical studies have been carried out on this model in the recent past, makes it a perfect test-case. For the sake of comparison with other numerical methods, and in order to ensure that our algorithm did not produce systematic errors, we also performed several tests on the two-dimensional model with nearest-neighbor (NN) interactions, for which exact results are known (see \cite{BorgsJanke1992,Buddenoir1993}; also references in \cite{Janke2002proc}). Both models are known to exhibit a first-order transition for an appropriate set of parameters, namely, $q>4$ for the NN model \cite{Wu1982}, and $\sigma < \sigma_c(q)$ for the LR one, with for instance $\sigma(3)=0.72(1)$ \cite{ReynalDiep2004a}. We chose a set of parameters that would allow us to observe both weak and strong first-order transitions, and concentrated on several indicators of performance, reliability, and scalability: these include tunneling, equilibrium and effective autocorrelation times, and mean acceptance rates.
These indicators inform us on the efficiency with which the Markov chain reaches the equilibrium distribution and explores the phase space. They also tell us at what rate successive measurements decorrelate from each other; hence what amount of resources is needed to obtain reliable statistics. Overall, they are therefore good indicators as to whether CPU resources are efficiently utilized or not. As regards scalability, we also computed the dynamical exponents associated with tunneling and equilibrium times; these indicate how fast needs in CPU time grow with the lattice size. 

All densities of states were calculated by means of the Wang-Landau algorithm, whereby, starting from an initial guess of the density of states $n(E)$, we update $n(E)$ after each visit to energy level $E$ according to the rule $\ln n(E) \leftarrow \ln n(E) + \ln f$, where $\ln f$ is hereafter termed \textit{Wang-Landau modification factor}.
In the case of LR models, the unequal spacing of energy levels and the existence of energy gaps in the vicinity of the ground state required that we introduced a few changes over the original version. In particular, using an interpolator for $\ln n(E)$ turned out to be mandatory in order to compensate for the finite width of histogram bins --- as would also be required for models having a continuous symmetry; indeed, we observed that using large bins tends to strongly reduce the acceptance rate if no interpolator is used. Bezier splines provide good interpolators, although a linear interpolation with a slope given by the microcanonical temperature $\beta(E)$ also proved to be particularly efficient whenever this last quantity was made available by other means, e.g., the transition matrix.

For small and medium lattice sizes, we systematically performed all simulations twice, first with standard single-spin updates (SSU) and then with our method embedding cluster updates (CU). We give an estimate of the error in the density of states obtained from both types of update schemes. For the largest lattice sizes we studied, however, the SSU implementation simply turned out to be impracticable, due to either exceedingly high tunneling times, and - for LR models -  excessive CPU demands, and we present results for the CU algorithm only. 
 
\subsection{Phase space exploration and mean acceptance rates}

\begin{figure}
	\centering
	\InsertFig{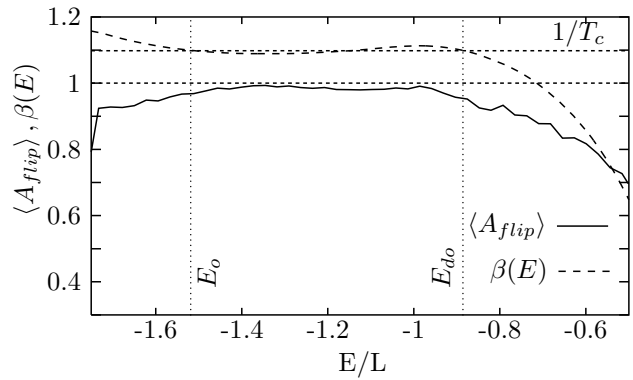}
	\caption{Mean acceptance rate as a function of the energy per spin for the six-state long-range Potts chain with $\sigma=0.7$, and $L=1024$ spins. The dashed line shows the estimated inverse microcanonical temperature.
	The vertical dotted lines indicate the position of the histogram peaks corresponding to the ordered and disordered phases.}\label{fig:mean_acceptance_rate}
\end{figure}

As opposed to the (canonical) Swendsen-Wang cluster algorithm, the acceptance rate of our algorithm --- Eq.~(\ref{eq:cluster_flip_acceptance_rate}) --- is not trivially equal to unity. Still, it is tightly related to the efficiency with which the Markov chain wanders about the phase space, since a low acceptance rate would lead to very repetitive dynamics. In this view, it is instructive to compute an approximate analytical expression of this acceptance rate when the initial and the final energies $E_a$ and $E_b$ differ only by a small amount. Writing $E_b = E_a + \epsilon$, and carrying out a series development to first order in $\epsilon$, one obtains $A_{\mbox{flip}} = \min \left(1, 1+ \Delta(E_a) dE\right)$, where 
$$
	\Delta(E_a) = \beta'(E_a) \left( \sum_{b_{ij}\in\mathcal{B}} J_{ij} \frac{1+p_{ij}(E_a)}{p_{ij}(E_a)} - |E_a| \right),
$$
with the same notation as in Sec.~\ref{sec:algorithm}.
We wish to obtain an estimate of the first statistical moments of $\Delta(E)$.
We hereafter consider the case of a model with nearest-neighbor interactions ($J=1$), for which we can carry out an exact derivation. The last expression simplifies to 
$$
	\Delta(E_a) = \beta'(E_a) \left(B \frac{1+p}{p} - |E_a| \right),
$$
where $B$ stands for the total number of bonds and $p=p(E_a)=e^{\beta(E_a)} -1$.
At a given energy $E_a$, at most $|E_a|$ bonds may be activated. The probability to activate a bond is $\pi(E_a) = p(E_a)/[1+p(E_a)] = 1 - e^{-\beta(E_a)}$; hence the probability to create a bond set $\mathcal{B}$ containing $B$ bonds writes $P(B)=\binom{|E_a|}{B} [\pi(E_a)]^B [1-\pi(E_a)]^{|E_a|-B}$, which may be reexpressed as 
$$
P(B) = \binom{|E_a|}{B} \frac{p^B}{(1+p)^{|E_a|}}.
$$
From there, we can derive the average bond count, $\meanval{B} = |E_a| \frac{p}{1+p}$. This allows us to rewrite $\Delta(E_a)$ as 
$$
\Delta(E_a)= \beta'(E_a) \frac{1+p}{p} ( B - \meanval{B});
$$
hence $\meanval{\Delta(E_a)}=0$. The variance of $\Delta(E_a)$ is thus proportional to the variance of the bond count distribution, i.e., $\meanval{B^2}-\meanval{B}^2=|E_a| \frac{p}{(1+p)^2}$, which yields
$$
	\sqrt{\meanval{\Delta(E_a)^2}} = \delta\Delta(E_a) = |\beta'(E_a)| \sqrt{\frac{|E_a|}{\exp{\beta(E_a)}-1}}
$$
For a given $\epsilon>0$, one half of all attempted cluster flips thus leads to an acceptance rate which is lower than $1$, the other half saturating at unity. Assuming a gaussian distribution for $\Delta(E)$, with the standard deviation computed above (which is valid for large enough lattice sizes), the mean acceptance rate is readily obtained from the mean value of a gaussian distribution centered at unity and truncated above $1$, which yields
$$
	\meanval{A_{\mbox{flip}}}(E_a) = 1 - \frac{\delta \Delta(E_a)}{2\sqrt{2\pi}} \epsilon
$$
In the case of interactions depending on  the distance $l$ between spins, one may observe that the average energy is related to the average number of bonds of length $l$ by $-\meanval{E} = \sum_{l>0} J(l) \frac{1+p_l}{p_l} \meanval{B(l)}$, which shows that $\meanval{\Delta(E)}=0$ also in this case.

At a first-order transition, $\beta(E)$ varies smoothly between the energy peaks of the ordered and disordered phases, which ensures that $\Delta(E)$ remains small.
The mean acceptance rate for the six-state LR Potts chain with $\sigma=0.7$ and $L=1024$ spins is sketched in Fig.~\ref{fig:mean_acceptance_rate}. While the acceptance rate is close to $1$ inside the range of phase coexistence, the variance of $\Delta(E)$ increases when $E$ lies outside the range of phase coexistence, and therefore leads to a reduction in the acceptance rate. We observe that this diminution is less marked at low-energy levels, for the energy cost associated with flipping a small number of big clusters is lower than that associated with randomly updating a great deal of small clusters, and $E_b-E_a$ is consequently lower in the last case. It is worth underlining, however, that the energy range of interest in the analysis of first-order phase transitions spans an interval which is only moderately larger than that corresponding to phase coexistence, the only requirement being that metastability plateaus \cite{ReynalDiep2004a} and histogram peaks must be clearly visible. As a result, the fact that the mean acceptance rate for cluster flips remains well above $90\%$
inside this range of energy represents already an improvement of a factor $3$ with respect to the standard multicanonical approach, where we obtained acceptance rates oscillating around $30\%$.

\subsection{Dynamic properties}

\begin{figure}
	\centering
	\InsertFig{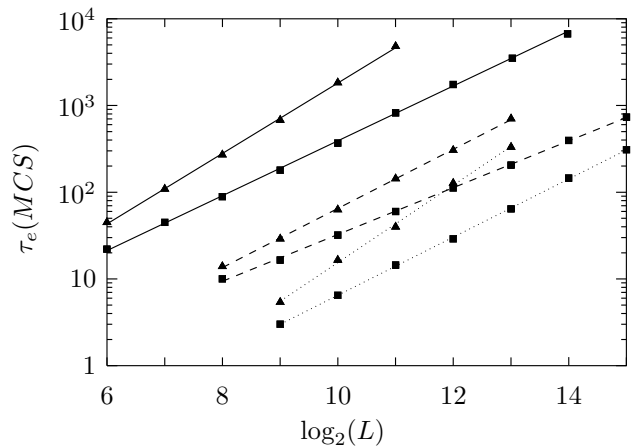}
	\caption{Tunneling times for the long-range Potts chain with $q=3$, $\sigma=0.4$ (dashed lines) and $0.6$ (dotted lines), and $q=6$, $\sigma=0.7$ (solid lines). Triangles refer to the SSU implementation, while squares indicates estimates for our method (CU). Dynamic exponents $z$ were determined from a fit to the power law $\tau_e \sim L^{z}$.}
	\label{fig:dyn_exp_potts1D}
\end{figure}

\begin{table}
	\caption{Dynamic exponents $z$ for the $q$-state Potts chain with power-law decaying interactions (a) and its two-dimensional counterpart with nearest-neighbor interactions (b). $z(SSU)$ and $z(CU)$ refer to single-spin and cluster updates respectively, while $z_{muBo}$ and $z_{muClus}$ make reference to the multibond method \cite{JankeKappler1995b} and Rummukainen's multi-microcanonical cluster method \cite{Rummukainen1993} applied to the NN model. }
	\begin{ruledtabular}
	  \begin{tabular}{llllll} 
	    $q$ 	&	$\sigma$	& 	$z(SSU)$	& 	$z(CU)$	&	$z_{muBo}$	& $z_{muClus}$ \\ \hline 	
		$6^a$	&	0.7			&	1.35(3)		&	1.05(1)		&	\\ 
		$3^a$	&	0.6			&	1.48(2)		&	1.11(1)		&	\\ 
		$3^a$	&	0.4			& 	1.13(2)		&	0.89(1)		&	\\ \hline
		$7^b$	&				&	2.60(4)		&	1.82(2)		&	1.84 & 1.82(3) \\ 
		$10^b$	&				&	2.87(4)		&	2.23(1)		&	2.1	\\ 
	  \end{tabular}
	\end{ruledtabular}
	\label{table:dyn_exp}
\end{table}

Where performance measurements at first-order transitions are concerned, tunneling times have thus far been regarded as one of the most meaningful measurement parameters \cite{JankeSauer1994,Janke2002proc,JankeKappler1995b}. They are defined as one half of the average number of MCS needed for the walk to travel from one peak of the energy histogram to the other -- where peaks are defined with respect to the finite-size transition temperature -- and turn out to represent a fairly good indicator of the interval between roughly independent samples.

Results for the LR chain with $q=3$ and $6$ are shown in Fig.~\ref{fig:dyn_exp_potts1D}.
Dynamic exponents $z$ were determined from a fit to the power law $\tau_e \sim L^{z}$, and are summarized in Table~\ref{table:dyn_exp}. We can witness a substantial reduction for both the LR and the NN models, with exponents close to and sometimes even below the ideal random-walk value $z=D$.
As regards the NN model, our values compare extremely well with those obtained with the multibond method \cite{JankeKappler1995b} and with Rummukainen's hybrid-like two-step algorithm \cite{Rummukainen1993}, although these approaches and ours differ markedly in the way clusters are constructed.

It should be mentioned that the distance $E_d-E_o$ the random walker must travel, i.e., the energy gap between the peaks of the histogram, does not scale linearly with the number of spins. This feature is especially apparent for long-range interactions, where $E_d-E_o$ grows all the more faster with increasing lattice size that $\sigma$ comes closer to $0$. As a result, the power law $\tau_e \sim L^{z}$ yields dynamical exponents which are underestimated with respect to the value given by a power law of the form $\tau_e \sim (E_d-E_o)^{z}$ (up to a dimensional factor $2$ for the NN model). For instance, we would obtain $z=1.40(3)$ instead of $z=1.35(3)$ for $q=6$ and $\sigma=0.7$, and $z=1.10(1)$ instead of $z=1.05(1)$. Where the performance in terms of CPU demands is concerned (and in particular if one is interested in how it grows with the size of the system), we think however that the traditional definition $\tau_e \sim L^{z}$ is more meaningful.

While tunneling times represent a practical way to estimate the efficiency with which the random walker drifts along the energy landscape, they are subject to two limitations. First, they cannot be properly defined in the case of second-order phase transitions, since the histogram of the energy does no longer exhibit two peaks. Second, there is no direct connection between tunneling times and autocorrelation times, which makes it difficult to estimate the optimum interval between measurements that will yield perfectly uncorrelated data, and thus minimum statistical error in estimates of thermodynamic data. It is worth mentioning here that computing integrated autocorrelation times naively from the set of measurements, i.e., just as is usually done in the canonical case, simply makes no sense when simulating in the multicanonical ensemble, because the quantities we are interested in are, in the first place, reweighted averages of thermodynamical data \cite{Janke2002proc}.

Therefore, alternate definitions have been proposed, which try to circumvent these limitations. One approach is to compute the so-called \textit{round-trip times} \cite{Alder2004}, which are computed from the number of MCS needed to get across the whole energy axis, that is, from the ground state to the upper energy level. Although round-trip times may be determined for any order of phase transition, they present unfortunately no more connection with statistical errors than do tunneling times. On the contrary, multicanonical \textit{effective autocorrelation times}, which were first introduced in the framework of the multibond algorithm \cite{JankeKappler1995b}, offer a direct comparison with exponential or integrated autocorrelation times of traditional use in canonical simulations. Mimicking the canonical case, the effective autocorrelation time $\tau_{\mbox{eff}}$ can be defined for any thermodynamic variable $\theta$ by inverting the standard error formula
$\epsilon_\theta^2 = \sigma_\theta^2 2 \tau_{\mbox{eff}}/N$, where $N$ stands for the total number of (possibly correlated) measurements, $\sigma_\theta^2$ denotes the variance of the (reweighted) thermodynamic variable $\theta$, e.g., $\meanval{E^2}-\meanval{E}^2$, and $\epsilon_\theta^2$ is the squared statistical error in the same variable. The error may be estimated either from resampling or (jackknife) blocking procedures, or by performing multiple independent runs. Since both the variance and the error depend on the reweighting temperature, the previous definition obviously yields an effective autocorrelation time which also depends on the temperature.    

\begin{figure}
	\centering
	\InsertFig{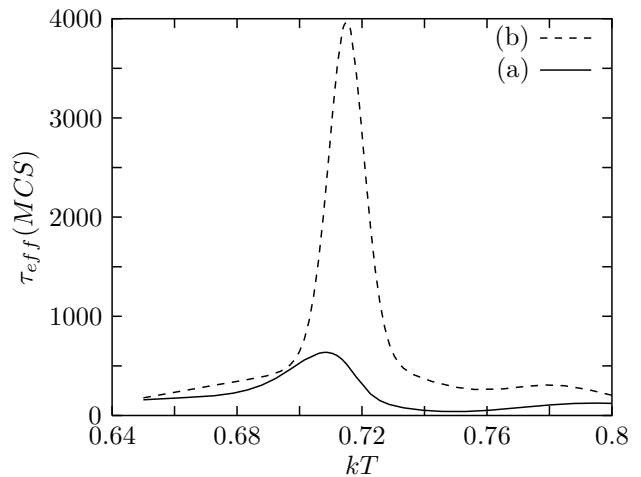}
	\caption{Effective autocorrelation time $\tau_{\mbox{eff}}$ for $q=6$, $\sigma=0.9$ and $L=512$ with (a) cluster updates (b) single-spin updates. The effective transition temperature defined from the peak of the specific heat is $T_c(C_v)=0.7163(2)$.}
	\label{fig:tau_eff_from_prod_runs_q6s9L9}
\end{figure}

We now discuss our results for effective autocorrelation times obtained for the six-state LR Potts chain with $\sigma=0.9$ and $128 \leq L \leq 1024$ spins. For this value of $\sigma$, the model exhibits a very weak first-order transitions with no clearly visible histogram peaks for sizes below $L \sim 2000$. The choice of medium lattice sizes was dictated by the fact that we computed the error from multiple independent runs (around 20 runs of $10^6$ MCS each), which we found a more reliable way of computing the statistical error than using a blocking procedure. Figure~\ref{fig:tau_eff_from_prod_runs_q6s9L9} shows the dependence of $\tau_{\mbox{eff}}$ on the temperature for $L=512$. For both algorithms, $\tau_{\mbox{eff}}$ exhibits a peak in the vicinity of the effective transition temperature $T_c(C_v)=0.7163(2)$. As expected, the reduction brought about by cluster updates in terms of correlation between measurements is marked, especially in the transition region, where single-spin update lead to a critical slowing down similar to the one encountered in canonical simulations. This behavior is consistent with the very general observation reported recently in \cite{Trebst2004} in the framework of the optimal ensemble, and also in \cite{Guerra2004} in the context of equilibration time for multicanonical algorithms (see also the next paragraph for more details on this issue), whereby the random walker diffuses at a slower pace in the critical region. In this respect, cluster updates optimize the diffusive current of the random walker in the critical region in much the same way as do the optimal ensemble weighting proposed in \cite{Trebst2004}, yet with a distinct strategy: in the latter, the error is reduced by allowing the walker to spend more time in the critical region than in the rest of the energy axis; in our approach, it is the decorrelating capability of the move update itself which reduces the statistical error in the transition region. As is well known, however, cluster updates are especially efficient at the percolating threshold, and the reduction in terms of correlation is large \textit{because} bond probabilities are governed by the microcanonical temperature. This interpretation is clearly underpinned by our investigation of the effect of poor estimates of $\beta(E)$ on tunneling times, presented later in Sec.~\ref{sec:estimating_betaE}. 
\begin{table}
\caption{Effective autocorrelation times at the transition temperature defined from the location of the peak of the specific heat, for the six-state LR Potts chain with $\sigma=0.9$}  
\begin{ruledtabular}
\begin{tabular}{lll} 
$L$ 	&	$\tau_{\mbox{eff}}(SSU)$	&	$\tau_{\mbox{eff}}(CU)$ 	\\ \hline
$128$	&	475					&	155					\\
$256$	&	1390				&	310					\\
$512$	&	3960				&	635					\\ 
$1024$	&	12700				&	1370				\\
$z$ &	1.6(1)				&	1.0(1)				\\
\end{tabular}
\end{ruledtabular}
\label{tab:tau_eff_vs_L_q6s9}
\end{table}
\begin{figure}
	\centering
	\InsertFig{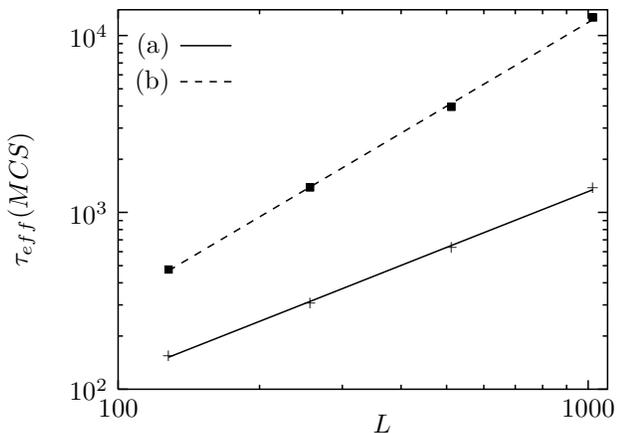}
	\caption{Fit of effective autocorrelation times $\tau_{\mbox{eff}}$ to the power law $\tau_{\mbox{eff}} \propto L^\alpha$ for the six-state Potts chain ($\sigma=0.9$ and $L=512$) with (a) cluster updates (b) single-spin updates.}
	\label{fig:tau_eff_vs_L_q6s9}
\end{figure}
Finally, we focus on the scaling behavior of autocorrelation times. Table~\ref{tab:tau_eff_vs_L_q6s9} reports our results for $L$ ranging from 128 to 1024 spins, where $\tau_{\mbox{eff}}$ is evaluated at the effective transition temperature determined from the peak of the specific heat. Our method gives smaller autocorrelation times already for $L=128$ spins. From these values, we also determined the associated scaling exponents by a fit to the power law $\tau_{\mbox{eff}} \propto L^z$ (Fig.~\ref{fig:tau_eff_vs_L_q6s9}), and obtained a highly satisfying value of $z \sim 1.0(1)$ with cluster updates. 

We conclude the discussion on the dynamic characteristics of our algorithm with an investigation of equilibrating properties. As opposed to canonical simulation, estimating equilibrium times has been much less common in the context of multicanonical simulations; the non-linear relaxation function, while very informative when the equilibrium distribution is driven by a Boltzman weight \cite{LandauBinderBook2000}, is of limited use indeed if the engendered distribution is flat.  Recently, however, an efficient procedure aimed at  estimating equilibrium times for any equilibrium distribution was proposed by Guerra and Mu\~noz \cite{Guerra2004}. This procedure relies on a $\chi^2$ regression with respect to the (expected) flat equilibrium distribution $\mathcal{P}(E)$. Starting from the same initial lattice configuration, $n$ Markov processes are run with distinct random seeds, and at each MC step $t$, a histogram of the energy $V_t(E)$ is filled with the value of the energy of each process. Asymptotically, $V_t(E)$ should approximate the expected flat distribution $\mathcal{P}(E) \propto n(E) w(E)$. In order to estimate the equilibrium time in a more quantitative way, a $\chi^2(t)$ deviation of $V_t(E)$ with respect to the flat distribution is carried out at each MC step $t$, i.e.,  $\chi^2(t) = \sum_E (V_t(E)-n\mathcal{P}(E))^2/(n\mathcal{P}(E))$, where the sum runs over histogram bins. For large $n$, and provided equilibrium has been reached, the distribution of $\chi^2(t)$ over $m$ experiments obeys a $\chi^2$ law with a number $r$ of degrees of freedom given by the number of histogram bins minus one, that is, with a mean equal to $r$ and a standard deviation given by $\sqrt{2r/m}$. 
\begin{figure}
	\centering
	\InsertFig{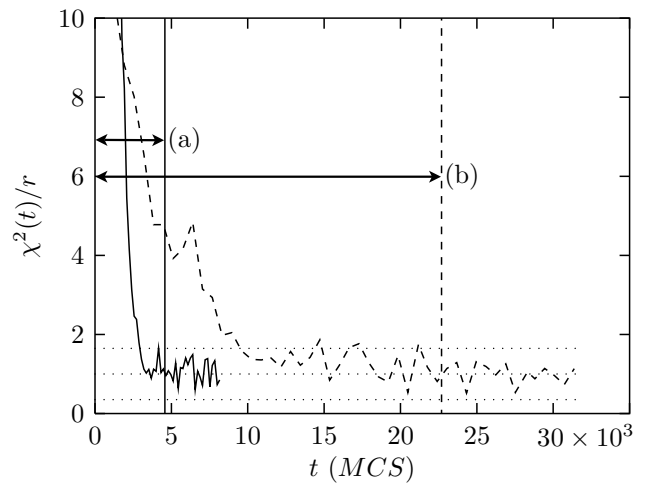}
	\caption{Plot of $\chi^2(t)/r$ for the six-state Potts chain ($\sigma=0.9$, $L=512$) using (a) cluster updates and (b) single-spin updates. The regression was carried out over a histogram containing $20$ bins populated from $1000$ runs, all starting in ground state configuration but with distinct random seeds.}
	\label{fig:equ_time_q6s9l512}
\end{figure}
Due to the intensive demand in CPU required by this procedure, we restricted our  estimation of equilibrium times to the single case $q=6$ and $\sigma=0.9$. We performed $n=1000$ Markov processes for sizes between $L=128$ and $L=512$, and estimated the equilibrium time from a single experiment (that is, $m=1$) by simply monitoring the time needed for $\chi^2(t)/r$ to reach unity and then stay within the interval $[1-2\sigma/r,1+2\sigma/r]$. As illustrated in Fig.~\ref{fig:equ_time_q6s9l512}, relying on a single experiment leads to quite large error bars, yet this is sufficient for our purpose. From the graphs of $\chi^2(t)$ we read  $\tau_{eq}=4500 \pm 500$ MCS and $\tau_{eq}=23000 \pm 2000$ MCS for the cluster- and single-spin updates respectively; in spite of the large uncertainty, the reduction in terms of equilibrium time brought about by our method is clearly visible. 
\begin{table}
\caption{Equilibrium times for the six-state LR Potts chain with $\sigma=0.9$ obtained by monitoring the graph of $\chi^2(t)/r$.}
\begin{ruledtabular}
\begin{tabular}{lll} 
$L$ 	&	$\tau_{eq}(SSU)$	&	$\tau_{eq}(CU)$ 	\\ \hline
$128$	&	1700(100)			&	800(120)			\\
$256$	&	6000(750)			&	2000(200)			\\
$512$	&	23000(2000)			&	4500(500)			\\
$1024$	&	101000(8000)		&	10000(800)			\\
\end{tabular}
\end{ruledtabular}
\label{table:tau_eq_vs_L_q6s9}
\end{table}
\begin{figure}
	\centering
	\InsertFig{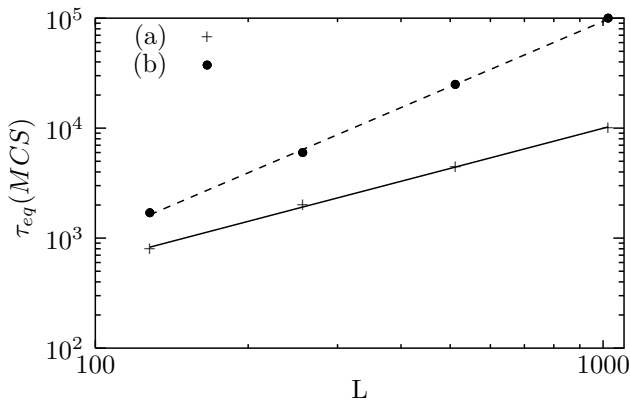}
	\caption{Fit of equilibrium times to the power law $\tau_{eq} \propto L^\alpha$ for the six-state Potts chain ($\sigma=0.9$). (a) cluster updates and (b) single-spin updates.}
	\label{fig:tau_eq_vs_L_q6s9}
\end{figure}
Results for other lattice sizes are summarized in Table~\ref{table:tau_eq_vs_L_q6s9}. A fit to the power law $\tau_{eq} \propto L^z_{eq}$ (see Fig.~\ref{fig:tau_eq_vs_L_q6s9}) yields the scaling exponents $z_{eq}=1.96(5)$ and $z_{eq}=1.21(3)$ for the single-spin and the cluster updates respectively. Here again, we think that lower diffusion currents in the latest case account for the higher pace at which the random walker reaches the equilibrium distribution.

\subsection{Overall CPU demand for LR models}

We now discuss CPU demand in the case of LR models, and concentrate on the gain in CPU resources brought about by the optimization schemes proposed in Sec.~\ref{sec:lr_optimization}.
Assuming a decently efficient algorithm implementation, this indicator yields a rough account of the real algorithm complexity, although it should be mentioned that it is usually an elaborate task to estimate this quantity rigorously, partly because its value hinges heavily on a variety of implementation, CPU architecture and compiler dependent properties. We decided to measure CPU times over a series of one-hour long simulation runs on a handful of distinct CPU architectures, including Intel Pentium and Xeon at 2.4 and 3.2GHz. \begin{figure}
	\centering
	\InsertFig{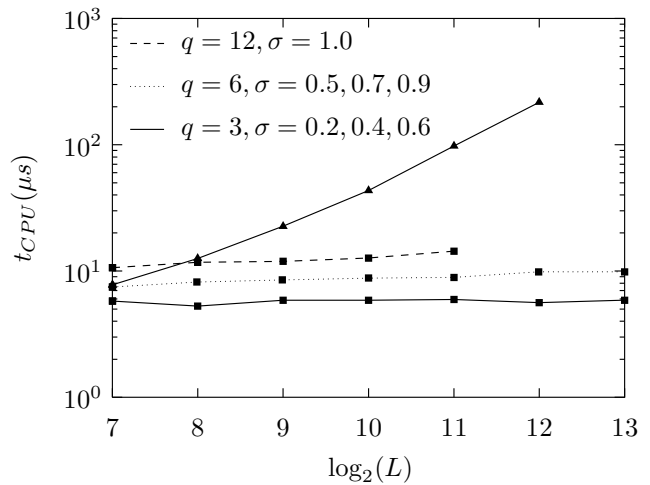}
	\caption{CPU time per MCS and per spin for the long-range Potts chain.
	Triangles indicate typical CPU times for the local-update algorithm (SSU), irrespective of $q$ and $\sigma$.
	Squares refer to our algorithm (CU) with LR specific optimizations included; for $q=3$ and $q=6$, estimates were determined by averaging over the indicated $\sigma$ values.}
	\label{fig:cpu_time_1D}
\end{figure}
Figure~\ref{fig:cpu_time_1D} sketches averages of the CPU (user) time per MCS and per spin, where small fluctuations might be attributed to the effect of varying CPU cache sizes amidst our clutch of CPU's. While for the local-update implementation the demand in CPU per spin grows linearly with the number of spin, it is roughly constant over a fairly large range of lattice sizes in the case of our cluster-update algorithm. Moreover, our method already outperforms the local-update scheme starting from several hundreds spins, with nonetheless an increased footprint for higher $q$ values which is accounted for by the correspondingly higher number of FFT's to be computed. This, however, clearly demonstrates the breakthrough that our method brings about for the study of long-range models, paving the way for precise tests of finite-size scaling.

\section{Two-dimensional NN Potts model: comparison with exact results}\label{sec:NN_model}

In order to check that our algorithm did not produce systematic errors, we computed transition temperatures and interface tensions for the two-dimensional $q$-state Potts model ($q=7,10$) with nearest-neighbor (NN) interactions and helical boundary conditions. Results regarding the dynamic characteristics of our algorithm for this model were reported in Sec.~\ref{sec:results_dyn_behaviour}; we will concentrate here on precision matters.
For $q=10$, we obtained $T_c(L)=0.70699(5)$, $0.70491(5)$, $0.70300(2)$, $0.70278(1)$, $0.70164(1)$, $0.701328(4)$ and $0.701249(2)$ for $L=16$, $20$, $30$, $32$, $64$, $128$, and $256$, where $T_c$ was determined from the location of peaks of the specific heat. $C_v$ was computed directly from the estimated density of states, and then refined from an additional production run of length $10^7$ MCS. The error was estimated by means of the jackknife method. Following standard finite-size scaling theory at first-order phase transitions, we collapsed $C_v(T)/L^2$ vs $(T-T_c)L^2$ over the five highest lattice sizes and found an infinite size temperature $T_c(\infty)=0.701236(3)$ in very good agreement with the exact value $0.70123157\ldots$
The same procedure applied to $q=7$ and $L=64$, $128$, and $256$ yielded $T_c(\infty)=0.773059(1)$ which again matches perfectly the exact value $0.7730589\ldots$
We estimated the interface tension (i.e., per unit length of the interface) $f^s$ between the ordered and disordered phases from the histogram of the energy, reweighted at a temperature where energy peaks have the same height \cite{Janke1992}: $2 f^s = -L^{-1}\ln P_{min}$, where the factor 2 arises from the use of periodic boundary conditions. Here, $P_{min}$ denotes the minimum of the histogram between the two energy peaks; peak heights are normalized to unity. We computed $f^s$ directly from the density of states, and estimated the error from the additional production run. In this respect, it should be noted that estimating interface tensions directly from the density of states generally yields values that lie below those computed from histograms collected during production runs.
Our algorithm allowed us to determine $f^s$ with a four-digit precision for sizes up to $L=256$ and nonetheless rather modest statistics.
For the seven-state model, we obtained $2 f^s=0.0336(6)$, $0.0294(1)$, $0.02631(8)$, and $ 0.02384(9)$ for $L=32$, $64$, $128$, and $256$; a linear fit of the form $\Sigma \sim \Sigma(\infty)+c/L$ \cite{LeeKosterlitz1991} performed over the three largest sizes (i.e., for $L$ above the disordered phase correlation length $\xi\sim 48$ \cite{Buddenoir1993}) yielded the infinite size value $0.02230(11)$, still above the exact value $0.020792$, yet closer to it than estimates reported in several previous studies \cite{Rummukainen1993,JankeKappler1995b,Janke2002proc}. 

\section{LR Potts chain: error estimates in thermodynamical data}\label{sec:LR_model}

In this section, we discuss the precision of our results for the $q$-state Potts chain with algebraically decaying interactions, i.e., $J(r)=1/r^{1+\sigma}$. Our purpose is twofold. First, we estimate the error in the density of states $n(E)$ obtained from the Wang-Landau algorithm, so that we can obtain a better insight into the benefit of our method with regards to the iterative calculation of $n(E)$. Second, we determine confidence intervals on reweighted averages computed from an additional production run. Since computing thermodynamic quantities from a production run does not require that the histogram be perfectly flat, nor that the estimate of the density of states be perfectly accurate, this reduces to estimating the gain in precision brought about by lower autocorrelation times. 

\subsection{Statistical error of the density of states}
In order to compare the error in the density of states produced by the single-spin update implementation and our method, we performed for each method a series of 12 independent simulations with the Wang-Landau algorithm, all starting with the same initial guess of the density of states. The model parameters were set to $q=6$, $\sigma=0.9$ and $L=512$. This choice of parameters guarantees that, in spite of the modest lattice size we consider, autocorrelation times differ by a sufficient amount between the single-spin updates method and our method, so that the benefit may be clearly interpreted in terms of decorrelating capabilities. The initial guess of $S(E)$ was scaled up from an estimate obtained at $L=256$, and the updating factor of the Wang-Landau algorithm was initially set to $\ln f=5$.
We did not make use of all improvements to the original Wang-Landau algorithm, as proposed by Zhou and Bhatt in \cite{Zhou2003}, since these would have partly overshadowed the gain produced solely by lower autocorrelation times.  Indeed, we mainly focused on the systematic error (rather than the whole statistical error) that may show up during the first iterations. It was shown in \cite{Zhou2003} that this systematic error results from the combination of a large $\ln f$ coefficient with the presence of strong correlations between adjacent binning. We thus simply relied on the original histogram flatness criterion to switch from one iteration to another, and divided $\ln f$ by the same amount (namely, $5$) after each iteration which passed the flatness check. We found out, however, that using the criterion in  \cite{Zhou2003} instead, that is, averaging $n(E)$ on multiple independent runs after each iteration, and switching to the next iteration only after a given number of entries was recorded in the histogram (see Eq.~(12) in \cite{Zhou2003}), led to markedly lower statistical errors.
\begin{figure}
	\centering
	\InsertFig{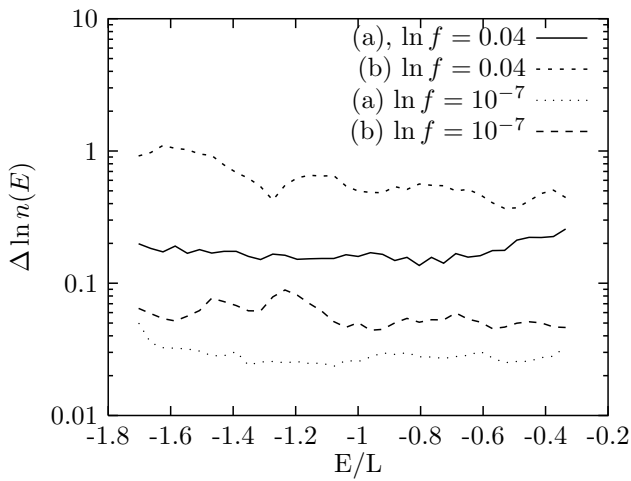}
	\caption{Statistical error in the density of states of the six-state Potts chain for two distinct modification factors $\ln f$ of the Wang-Landau algorithm. The statistical errors were obtained from $12$ independent runs. The parameters of the model are $\sigma=0.9$ and $L=512$. (a) and (b) correspond to our method and the local-update algorithm respectively.} 
	\label{fig:std_dev_SE_q6s9L9}
\end{figure}
As illustrated in Fig.~\ref{fig:std_dev_SE_q6s9L9}, the statistical error in the density of states is clearly improved by our method. In particular, cluster updates lead to a spread of the error over the whole energy axis. In this respect, and as already mentioned in Sec.~\ref{sec:results_dyn_behaviour}, the lower diffusion rates associated with collective updates in the critical energy region offer a clear benefit. As expected from the arguments of Zhou and Bhatt, the reduction is also more marked for $\ln f=0.04$ than for $\ln f=10^{-7}$, and the systematic error brought about by correlations between successive binning is indeed partly tamed by a lower Wang-Landau modification factor. 
\begin{figure}
	\centering
	\InsertFig{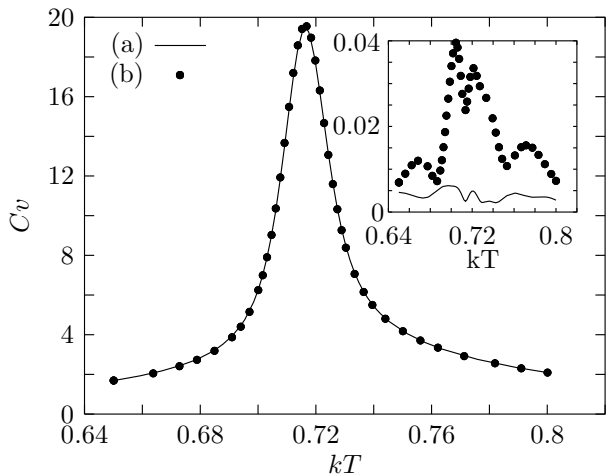}
	\caption{Graph of the specific heat for the six-state Potts chain ($\sigma=0.9$ and $L=512$) obtained directly from the final estimate of the density of states with (a) our method and (b) the local-update algorithm.
	The inset shows the relative error $\err(C_v)/C_v$.}
	\label{fig:errCv_over_Cv_q6s9L9}
\end{figure}
Finally, we show in Fig.~\ref{fig:errCv_over_Cv_q6s9L9} the resulting statistical error in the specific heat, since thermodynamical averages are the relevant quantities in the first place. $C_v$ was computed directly from the estimated density of states $n(E)$, i.e., according to the formula
$C_v(kT) = (\meanval{E^2}_{kT}-\meanval{E}_{kT}^2)/(L\ kT^2)$, where $\meanval{E^n}=(\sum_E E^n n(E) e^{-E/kT})/(\sum_E n(E) e^{-E/kT})$. For long-range models, energy levels are not equally spaced, and it should be noted that too large histogram bins may cause a systematic deviation of the averages as well. We paid attention to this by comparing our results for several bin widths, and made sure that the systematic deviation engendered was always lower than the statistical error itself. As shown in the inset of Fig.~\ref{fig:errCv_over_Cv_q6s9L9}, the accuracy of the estimation of $C_v$ is larger by nearly an order of magnitude at the transition temperature. Incidentally, we observe that this is comparable to the gain in terms of autocorrelation times, as already presented in Fig.~\ref{fig:tau_eff_from_prod_runs_q6s9L9}.

\subsection{Transition temperatures, Binder cumulants and interface free energies}

\begin{figure}
	\centering
	\InsertFig{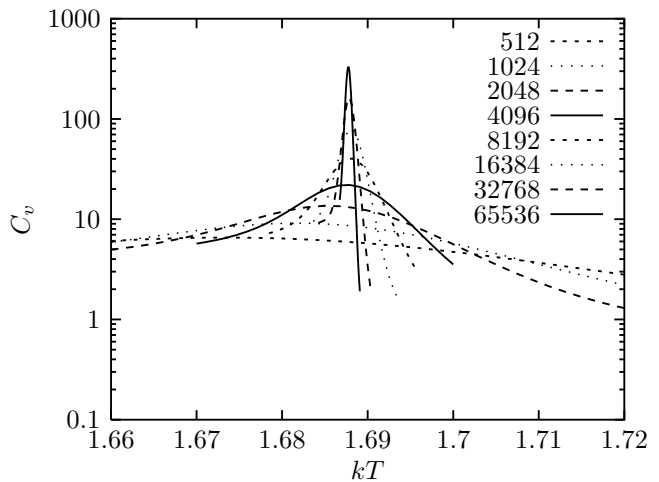}
	\caption{Specific heat for the three-state Potts chain with $\sigma=0.5$ as obtained with our method.}
	\label{fig:cv_1D_q3s5}
\end{figure}
\begin{table}
	\caption{Estimates of peaks of the specific heat $C_v$ and the susceptibility $\chi$, and corresponding effective transition temperatures for the three-state LR Potts chain with $\sigma=0.5$. Error calculations were carried out by means of the jackknife method applied to a single production run. The number of MCS per production run is the same for both methods, yet varies between $10^6$ and $10^7$ from the smaller to the larger lattice sizes. 
	}
	\begin{ruledtabular}
	  \begin{tabular}{lllll}
$L$		& $T_c(C_v)$	& 				& $C_v^{max}$	&			\\
	    & (CU) 			& (SSU) 		& (CU) 			& (SSU)		\\
128		& 1.6450(18)	& 1.645(3)		& 3.55(2)		& 3.55(3)	\\
256		& 1.6607(2)		& 1.6607(13)	& 4.86(2)		& 4.88(5)	\\
512		& 1.6741(9)		& 1.675(1)		& 6.54(3)		& 6.47(6)		\\
1024	& 1.6815(2)		& 1.6815(17)	& 9.14(8)		& 9.10(24) 	\\
2048	& 1.6856(3)		& 1.685(1)		& 13.63(15)		& 13.73(80)	\\
4096	& 1.68742(9)	& 1.6875(10)	& 22.21(34)		& 21.9(2.2)	\\
8192	& 1.68801(7)	&				& 40.28(44)		&		\\
16384	& 1.688031(34)	&				& 79.46(35)		&		\\
32768	& 1.687851(12)	&				& 164.1(4)		&		\\
65536 	& 1.687749(09) 	& 				& 332.8(6)		&		\\
$L$		& $T_c(\chi)$	& 				& $\chi^{max}$	&			\\
	    & (CU) 			& (SSU) 		& (CU) 		& (SSU)		\\
128		& 1.6793(14)	& 1.679(4)		& 3.44(3)		& 3.46(3)	\\
256		& 1.6837(2)		& 1.6837(15)	& 6.09(3)		& 6.13(5)	\\
512		& 1.6864(8)		& 1.6877(15)	& 10.81(7)		& 10.86(13)		\\
1024	& 1.6882(3)		& 1.688(2)		& 19.73(28)		& 19.8(5)	\\
2048	& 1.6887(2)		& 1.6882(15)	& 37.6(5)		& 37.5(1.8) 	\\
4096	& 1.68869(9)	& 1.6887(11)	& 75.4(1.2)		& 74.6(6.7)	\\
8192	& 1.68842(7)	&				& 165.2(1.8)	&	\\
16384	& 1.688148(35)	&				& 369.8(1.1)	&	\\
32768	& 1.687870(20)	&				& 827(2)		&	\\
65536 	& 1.687773(16)	&				& 1754(3) 		&	\\
	  \end{tabular}
	\end{ruledtabular}
	\label{table:Tc_q3s5}
\end{table}

We now discuss some of our results for the three-state Potts chain, for which we performed extensive simulations for sizes ranging from $L=128$ to $L=65536$. As opposed to higher values of $q$, there exists indeed a large collection of numerical studies for $q=3$, so that comparison with previous estimates is easier. Table~\ref{table:Tc_q3s5} reports our values for transition temperatures and peaks of response functions for $\sigma=0.5$. Both the specific heat $C_v$ and the susceptibility $\chi$ were computed from production runs whose length varied between $10^6$ and $10^8$ MCS depending on the lattice size, and error bars were computed by means of the jackknife blocking method. We performed these production runs twice, first using single-spin updates, and then using our method, yet in both cases with the same estimate of the density of states. Figure~\ref{fig:cv_1D_q3s5} shows the graph of $C_v$ as obtained with cluster updates. We mention that, for $L>4096$, the local-update implementation was simply intractable as a result of excessive computation times. For all sizes, our results match within error bars for both methods, and it should be noticed that, for the two largest sizes, we obtain estimates of finite-size transition temperatures accurate up to the fifth digit. 

\begin{figure}
	\centering
	\InsertFig{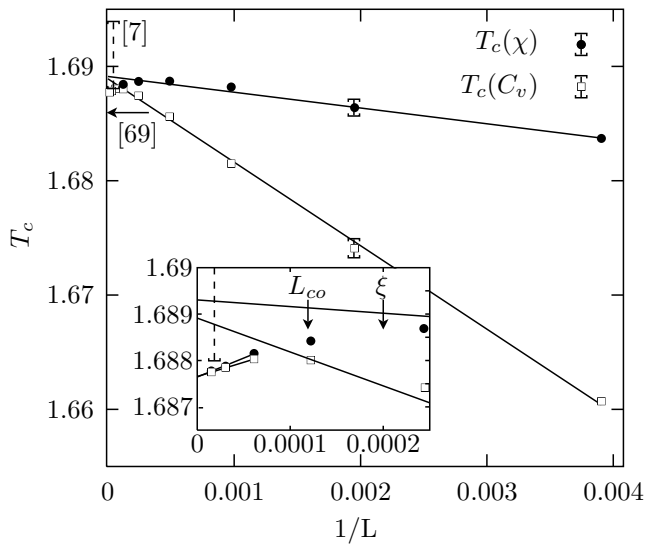}
	\caption{Solid lines show linear fits of finite-size transition temperatures vs $1/L$ for the three-state LR Potts chain with $\sigma=0.5$, performed over sizes ranging from $L=256$ to $L=4096$.  $T_c(C_v)$ and $T_c(\chi)$ are defined from the peaks of the specific heat $C_v$ and the susceptibility $\chi$ respectively; all data were obtained using cluster updates (CU). 
	The inset shows a detailed view at larger lattice sizes, along with a linear fit carried out over the three largest sizes $L=16384$, $L=32768$ and $L=65536$.
	Error bars are smaller than the size of symbols, except where explicitely indicated.
	The horizontal arrow shows the infinite-size estimate $T_c=1.68542$ obtained by Monroe using the cluster mean-field method \cite{Monroe1999}, while the vertical dashed error bar refers to the infinite-size estimate $T_c = 1.691(3)$ obtained in \cite{ReynalDiep2004a}. Finally, $L_{co}$ indicates the lattice size at which the graph of the peaks of $U_L$ vs the lattice size changes slope (see Fig.~\ref{fig:ULmax_vs_L_q3s2s5}), and $\xi$ refers to our estimate of the correlation length.
	}
	\label{fig:Tc_q3s5}
\end{figure}

In order to determine infinite-size transition temperatures, we performed a fit of the finite-size temperatures $T_c(L)$ reported in Table~\ref{table:Tc_q3s5}, to the power law $T_c(L)-T_c(\infty) = a/L$. This is illustrated in Fig.~\ref{fig:Tc_q3s5} where for the sake of clarity, only data obtained using cluster updates are reported.
An important effect we noticed is the presence of a crossover around $L=16384$ for $T_c(C_v)$, where the finite-size transition temperature reaches a maximum; a similar behavior can be witnessed for $T_c(\chi)$ with a maximum occuring around $L=4096$. In view of this, we first performed a fit over lattice sizes ranging from $L=256$ to $L=4096$: for this range of lattice sizes, data points lie neatly on a straight line within error bars for both $T_c(\chi)$ and $T_c(C_v)$. 
As regards the peaks of the specific heat, a fit performed over the range $L=[256-8192]$ yielded the same estimate of the infinite size temperature, which seems to indicate that $L=8192$ still belongs to the region where a linear fit is reliable.
Infinite size temperatures are reported in Table~\ref{table:Tc_q3s5_inf}: the temperatures computed from both methods compare very well with each other and with the value of $1.691(3)$ reported in \cite{ReynalDiep2004a} using a multicanonical approach and medium lattice sizes. The value obtained from a previous MC study based on a canonical version of the Luijten-Bl\"ote algorithm \cite{GlumacUzelac1997_1998a} lies clearly above our estimate, while estimates obtained from a transfer matrix method \cite{GlumacUzelac1993} and a real-space renormalization group approach \cite{CannasMagalhaes1997} fall markedly below our values. 
Finally, the best estimate determined so far (to the best of our knowledge) with a numerical approach, namely, the value of $T_c=1.68542$ obtained in \cite{Monroe1999} with the cluster mean-field method, lies slightly below ours. 
The occurence of a crossover for both response functions led us to perform a distinct linear fit restricted to the three largest lattice sizes (see the inset of Fig.~\ref{fig:Tc_q3s5} and  results in Table ~\ref{table:Tc_q3s5_inf}): this yielded $T_c(\infty)= 1.68764(1)$ and $T_c(\infty)= 1.68765(2)$ for $T_c(C_v)$ and $T_c(\chi)$ respectively. In terms of the number of digits of precision, these estimates are comparable to the value reported in \cite{Monroe1999}, although they still lie above it by around 0.13\%. 
This is all the more surprising that, even if the accuracy of the cluster mean-field method decreases with increasing $\sigma$ (with $\sigma \leq 0.5$ being quoted by the authors in \cite{Monroe1999} as the interval where the method produces its most accurate estimates), this method always produces --- by construction --- upper bounds for  transition temperatures, irrespective of the size of the clusters. 
Noteworthy enough, a similar discrepancy was reported by the authors of an MC study of the LR Ising model based on the Luijten-Bl\"ote algorithm \cite{LuijtenBlote1997}, with the cluster mean-field approach \cite{Monroe1998} yielding values underestimated by 0.8\% for $\sigma=0.5$.
Since the authors in \cite{Monroe1998,Monroe1999} performed an extrapolation procedure to compute the infinite-size estimate from temperatures obtained at finite cluster size, the discrepancy may thus be attributed to their specific extrapolation procedure rather than to the method used to produce finite-size estimates.

\begin{figure}
	\centering
	\InsertFig{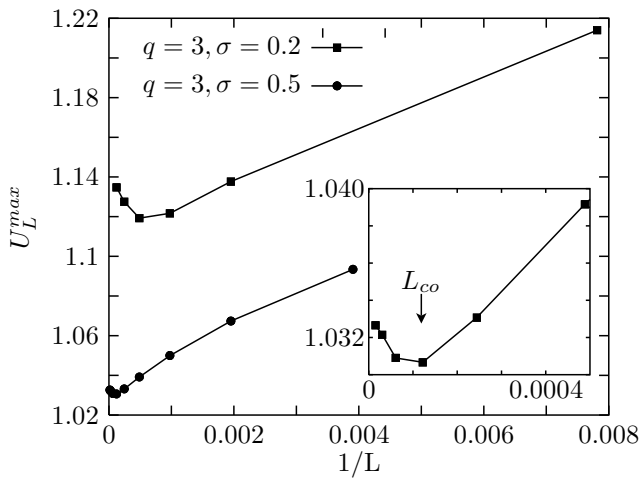}
	\caption{Peak of the cumulant of the energy $U_L=\meanval{E^4}/\meanval{E^2}^2$ as a function of the lattice size for the three-state LR Potts chain with $\sigma=0.2$ and $\sigma=0.5$. The inset shows a magnification near the origin for $\sigma=0.5$; $L_{co}$ indicates the approximate lattice size at which the curve changes slope, crossing over to a first-order transition behavior.} 
	\label{fig:ULmax_vs_L_q3s2s5}
\end{figure}

\begin{table}
	\caption{Infinite size transition temperatures computed from a fit of finite-size temperatures to a power law of the form $T_c(L)-T_c(\infty) = a/L$. The first column indicates the range of lattice sizes that were used in the fit. Results from previous studies are shown for comparison in the last column:
	Ref. \cite{ReynalDiep2004a} MC study based on Berg's multicanonical method and sizes up to $L=400$;
	Ref. \cite{Monroe1999} cluster mean-field method;
	Ref. \cite{GlumacUzelac1993} transfer matrix method;
	Ref. \cite{GlumacUzelac1997_1998a} MC study based on the Lujiten-Bl\"ote cluster algorithm and sizes up to $L=3000$;
	Ref. \cite{CannasMagalhaes1997} real-space renormalization group approach.
	}
	\begin{ruledtabular}
	  \begin{tabular}{llll}
$[L_1-L_2]$ & $T_c(C_v)$	& 			&	\\
			& (CU) 			& (SSU)		&	\\ 	
256-4096	& 1.6889(1)		& 1.6888(8)	&	\\
256-8192	& 1.6889(1)		&			&	\\
16384-65536 & 1.68764(1)	&			&	\\
			& $T_c(\chi)$	& 			&	\\
			& (CU)			& (SSU) 	&	\\
256-4096 	& 1.6893(1)		& 1.6892(6)	&	\\
16384-65536 & 1.68765(2)	&			&	\\
Ref. \cite{ReynalDiep2004a} &				&			& 1.691(3) 	\\ 
Ref. \cite{Monroe1999}	&				&				& 1.68542	\\
Ref. \cite{GlumacUzelac1993} &				&			& 1.664 	\\ 
Ref. \cite{GlumacUzelac1997_1998a} &		&			& 1.72  	\\ 
Ref. \cite{CannasMagalhaes1997}	&			&			& 1.41 		\\ 
	\end{tabular}
	\end{ruledtabular}
	\label{table:Tc_q3s5_inf}
\end{table}

In order to shed light on the presence of a crossover, we compared the lattice size at which the crossover occurs with two characteristic lengths. 
One such length is the lattice size $L_{co}$ where the peak of the reduced Binder cumulant of the energy, namely, $U_L^{max}=\max(\meanval{E^4}/\meanval{E^2}^2)$, experiences a minimum. At a second-order phase transition, $U_L^{max}$ tends trivially to $1$ in the thermodynamic limit, while it tends to a distinct value if the transition is of the first order. As illustrated in Fig.~\ref{fig:ULmax_vs_L_q3s2s5}, the graph of $U_L^{max}$ with respect to the lattice size is clearly characteristic of a first-order transition, with an infinite-size value lying close to $1.033$. A minimum occurs at $L_{co} \sim 8192$ and --- as shown in the inset of Fig.~\ref{fig:Tc_q3s5} --- the crossover occurs slightly above $L_{co}$ for $T_c(C_v)$, and slightly below for $T_c(\chi)$. For $\sigma=0.2$, the same correlation is witnessed by our results, with the change of slope of $U_L^{max}$ taking place around $L_{co}=2048$, and a change of behavior for $T_c(C_v)$ occuring near $L=4096$. 
Another characteristic length is the correlation length of the disordered phase, which we compute in a subsequent part of this section: for $\sigma=0.5$, our estimate of $\xi_d \sim 5000$ lies, here again, very close to the change of behavior of $T_c(C_v)$ and $T_c(\chi)$ (see the inset of Fig.~\ref{fig:Tc_q3s5}); for $\sigma=0.2$, $\xi_d \sim 600$, and $T_c(C_v)$ and $T_c(\chi)$ depart from the straight line at $L \sim 3000$ and $L \sim 1000$ respectively. For both values of $\sigma$, we thus observe the same qualitative behavior, namely, the crossover occurs at sizes slightly above $L_{co}$ and $\xi$. We think that this may be attributed for one part to the fact that, as long as the lattice size is smaller than the correlation length, the finite-size scaling behavior of the model is that of a continuous transition, whereas it is first-order-like at larger sizes. Another important finite-size effect may be specifically linked to the long-range nature of the interaction:  in \cite{ReynalDiep2004a}, it was suggested that, irrespective of the type of periodic boundary conditions implemented, the finite size of the lattice {\em shifts} the (effective) decay parameter $\sigma$ towards the mean-field regime, in a way that is more pronounced at small lattice size; this effect may thus compete with the previous effect in a non-trivial manner. It is therefore apparent that the power law used in our fits is not sufficient to model the finite-size scaling behavior in an accurate way over the entire range of lattice sizes; the derivation of an appropriate finite-size scaling law including corrections to scaling is left, however, to a subsequent work.

We also note in passing that, for $\sigma=0.5$, relying on the Binder cumulant to assess the first-order nature of the transition requires simulating the system up to sizes that are far beyond the capabilities of single-spin update implementations. In particular, performing a power-law fit of $U_L$ restricted to sizes below $L \sim 3000$ would yield underestimated values. Our results in Fig.~\ref{fig:ULmax_vs_L_q3s2s5} show that the infinite size value lies around $1.033$, and thus that the transition is stronger than suggested for instance in \cite{GlumacUzelac1997_1998a}.

\begin{figure}
	\centering
	\InsertFig{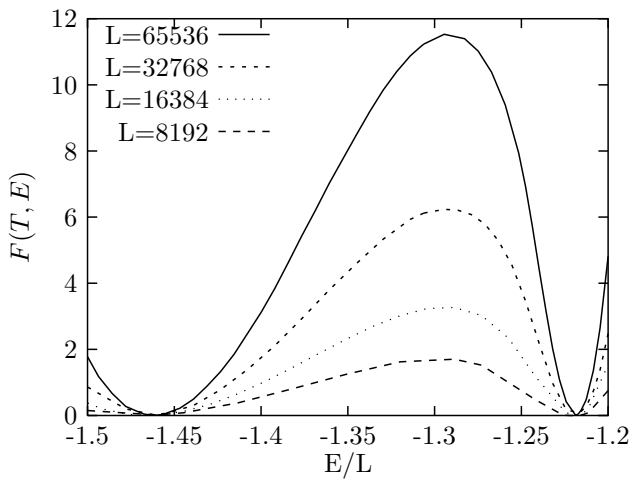}
	\caption{Graph of the free energy $F(T,E)=-\ln N(T,E)$ for the three-state LR Potts chain with $\sigma=0.5$. $N(T,E)$ is the histogram of the energy reweighted at a pseudo-transition temperature $T_{eqh}$ where both peaks have the same height.  For the four lattice sizes shown here, lattice configurations corresponding to phase coexistence are suppressed by a factor ranging from $0.1$ to $10^{-6}$ with respect to pure phase configurations; for the three largest sizes, we note that a canonical simulation is clearly intractable.}
	\label{fig:logH_q3s5}
\end{figure}
\begin{figure}
	\centering
	\InsertFig{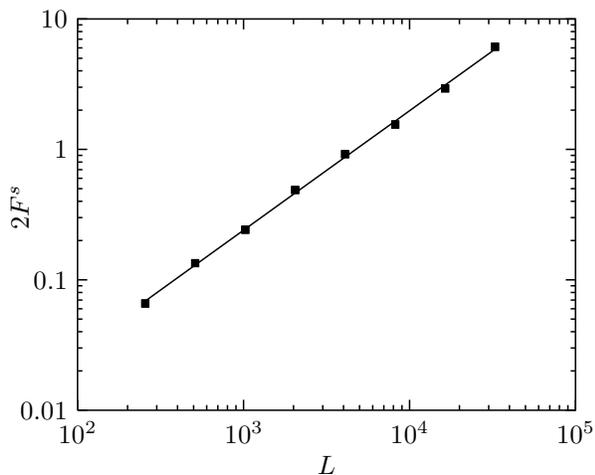}
	\caption{Fit of the interface free energy $F^s$ to a power law of the lattice size for the three-state LR Potts chain with $\sigma=0.5$. All estimates of $F^s$ were obtained with our method.}
	\label{fig:fit_dF_vs_L_q3s5}
\end{figure}
Although a precise determination of correlation lengths for the LR Potts chain is beyond the scope of this work, we tried to obtain rough estimates of them from the finite-size scaling behavior of the interface free energy $F^s$.
As in Sec.~\ref{sec:NN_model}, we computed a histogram $N(T,E)$ of the energy reweighted at the pseudo-transition temperature $T_{eqh}$ where both peaks of the histogram have equal height (see Fig~\ref{fig:logH_q3s5}). After normalization of the peak heights to unity, $2 F^s$ is given by $-\ln P_{min}$, where $P_{min}$ stands for the minimum of the histogram between the two energy peaks. With this definition, which is analog to that used in models with nearest-neighbor interactions, $2 F^s$ corresponds to the free energy cost associated --- when periodic boundary conditions are used --- with the creation of mixed ordered-disordered configurations from pure phase configurations. As we will witness, there is strong evidence that the dimension of the interface is no longer an integer.
By fitting the interface free energy to the power law $F^s \propto L^\alpha$, we obtain a very good fit for sizes ranging from $L=256$ to $L=65536$, yielding $\alpha=0.91(2)$ and $2F^s/L^\alpha=0.0004$ in the thermodynamic limit. This is illustrated in Fig.~\ref{fig:fit_dF_vs_L_q3s5}. For $\sigma=0.2$, we obtain $\alpha=0.74(3)$ from a fit performed over sizes ranging from $L=512$ to $L=8192$. In view of the expected behavior for nearest-neighbor interactions, namely, $F^s$ scales to leading order as a power of the lattice size with an exponent given by the dimension of the interface \cite{LeeKosterlitz1990}, this suggests that the effective dimension of the interface lies between $0$ and $1$ for long-range chains. 
This assumption is also supported by the fact that the fits of $F^s/L$ in \cite{GlumacUzelac1997_1998a} exhibit important finite-size corrections, while our fit with a non-integer exponent does not suggest such corrections. 

Finally, we can estimate the correlation length by transposing the argument that was proposed in \cite{BorgsJanke1992} to relate the interface tension $f^s$ of the nearest-neighbor Potts model to the correlation length $\xi_d$ of its disordered phase. This argument relies on two ingredients: 
(i) the correspondance between the order-order interface tension and the correlation length of the disordered phase, which results from duality; 
(ii) the assumption of complete wetting, which implies a simple relation between the order-order and order-disorder interface tensions, where the latter is the quantity $f^s$ we measure from the reweighted histogram of the energy. 
The second assumption is rigorous if $q$ is large enough, although the authors in \cite{BorgsJanke1992} suggest that it should stay valid for all $q$ for which the transition is of the first order. 
The argument leads to the relation $\xi_d^{-1}=2f^s=2F^s/L$ for the nearest-neighbor model: 
in view of the finite-size scaling behavior of $F^s$ reported above for the LR chain, it seems natural to transpose this relation by defining the interface tension as $f^s = F^s/L^\alpha$ for LR chains, and dimensional analysis thus suggests to use the relation $\xi_d^{-\alpha}=2f^s=2F^s/L^\alpha$ in the LR case.
This yields an estimate of $\xi_d \sim 5000$ for the LR chain at $\sigma=0.5$, a value that seems at least consistent with the fact that the change of slope of $U_L^{max}$ occurs at a lattice size slightly above this size (see Fig.~\ref{fig:ULmax_vs_L_q3s2s5}). For $\sigma=0.2$, the same relation yields $\xi_d \sim 600$, and here again this estimate is consistent with the behavior of $U_L^{max}$; in addition, and as expected, $\xi_d$ decreases with $\sigma$, i.e., as the transition becomes stronger.
Although there is no rigorous proof that the ingredients invoked in \cite{BorgsJanke1992} retain their validity in the case of LR interactions, it seems that a straighforward transposition to LR interactions yields reasonable results, even though the topology of the interface between the ordered and disordered phases is certainly far more complex than in nearest-neighbor models. In addition, we make use of Infinite Image Periodic Boundary Conditions \cite{CannasLapilliStariolo2004}, so that the factor $2$ in the definition of the interface tension might be questionable in LR models. Altogether our estimates should thus be taken as very rough ones.

\section{Combination with the transition matrix method} \label{sec:estimating_betaE}

In this section, we examine how our method can be efficiently combined with the transition matrix method \cite{Wang1999}. We show in particular that transition matrices represent a very efficient way of estimating the microcanonical temperature $\beta(E)$ used to compute cluster bond probabilities when nothing is known initially about the density of states. We also discuss how the estimated $\beta(E)$ can then be used as an efficient predictor to speed up the convergence towards the ground state during the early iterations of the Wang-Landau algorithm.

\subsection{Efficient estimation of $\beta(E)$ and bootstrapping}
As seems obvious from the scheme presented in Sec.~\ref{sec:algorithm}, one of the basic requirements of our algorithm is to have an estimate of $\beta(E)$ at our disposal over the whole energy axis  in order to compute cluster bond probabilities.   
\begin{figure}
	\centering
	\InsertFig{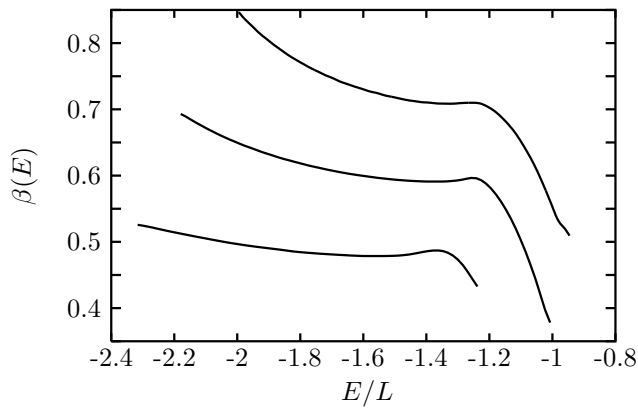}
	\caption{Microcanonical inverse temperature $\beta(E)=dS(E)/dE$ computed from the estimated density of states using a spline interpolation, 
	for the three-state long-range chain with $\sigma=0.4$, $0.5$, and $0.6$ from bottom to top.}
	\label{fig:betaE_from_dSdE}
\end{figure}
One rather simple way of estimating $\beta(E)$ is to compute it from the current estimate of the density of states $n(E)$ using a finite-difference scheme, i.e., in real-time in the course of the iteration scheme. This is the most tractable approach if one decides to rely solely on Wang-Landau's algorithm to estimate $n(E)$. During early iterations, however, the estimate of $n(E)$ is somewhat rough and it is necessary to resort to a spline interpolation in order to obtain a sufficiently smooth estimate of $\beta(E)$. 
Since the unequal spacing of energy levels in long-range models renders an interpolation scheme for $n(E)$ absolutely mandatory \cite{ReynalDiep2004a}, $\beta(E)$ is already available to us for free. Figure~\ref{fig:betaE_from_dSdE} shows estimates obtained with this approach for the three-state long-range chain with various interaction ranges, computed after ten iteration steps of $10000$ measurements each. We note in passing that the presence of a clearly visible minimum in the three cases results from the first-order nature of the transition. For sufficiently short-range interactions, and when no random disorder is present, the microcanonical entropy $S(E)$ scales quite gently with the lattice size, and it is also perfectly feasible to use the value of $\beta(E)$ obtained at a smaller lattice size as an initial guess.

\begin{figure}
	\centering
	\InsertFig{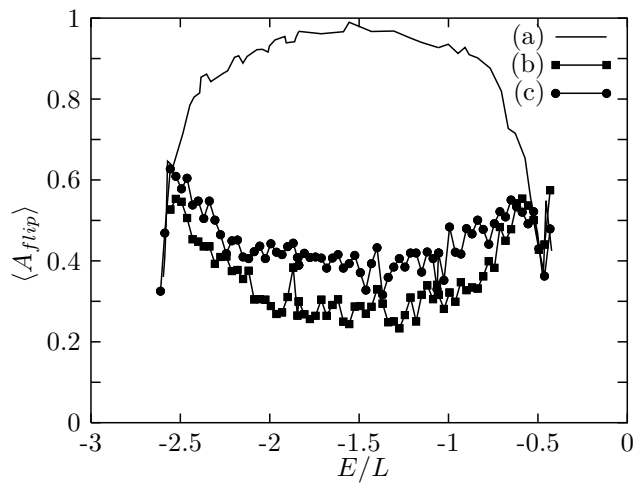}
	\caption{Mean acceptance rate as a function of the energy per spin for the six-state long-range Potts chain with $\sigma=0.5$, and $L=512$ spins (strong first-order regime) for three distinct estimates of $\beta(E)$. (a) best estimate, as given by the ultimate iteration of the Wang-Landau algorithm; (b) $\beta(E)$ scaled by $0.9$; (c) $\beta(E)$ scaled by $1.1$.}
	\label{fig:acceptance_rate_vs_error_betaE}
\end{figure}
In any case, it is crucial for the performance of our algorithm that we should compute $\beta(E)$ to sufficient accuracy. Indeed, we have found that any departure from the ideal line results in poorer performance, as illustrated in Fig.~\ref{fig:acceptance_rate_vs_error_betaE}. The curve (a) in the figure shows the mean acceptance rate as a function of the energy for an estimate of $\beta(E)$ obtained after the ultimate Wang-Landau iteration and a modification factor $\ln f=10^{-7}$. Curves (b) and (c) show the same quantity for microcanonical temperatures that were under- and overestimated by 10\%. 
The poor estimate of $\beta(E)$ causes a marked decrease of the acceptance rate in the transition region (around $E/L \sim -1.5$), from around 100\%
to nearly 40\%. 
Tunneling times obviously experience a corresponding increase, from $243$ for the best estimate, to $737$ and $1150$ for the under- and overestimated temperatures, respectively. This can be easily explained, if one considers that the efficiency of cluster updates reaches a maximum at the percolation threshold. Any departure of the estimate of $\beta(E)$ from the ideal line results in a shift between the temperature at which clusters percolate (which depends on $\beta(E)$) and the \textit{effective} temperature of the system (which is given by $dS(E)/dE$). This behavior has been observed in the context of canonical simulations of disordered systems, e.g., the Random Field Ising model \cite{Newman1996}, where the presence of randomness depresses the critical temperature. In this case, using the (canonical) simulation temperature to compute the bond probabilities simply results in a growing shift between the critical temperature and the percolation threshold as the randomness is increased.

\begin{figure}
	\centering
	\InsertFig{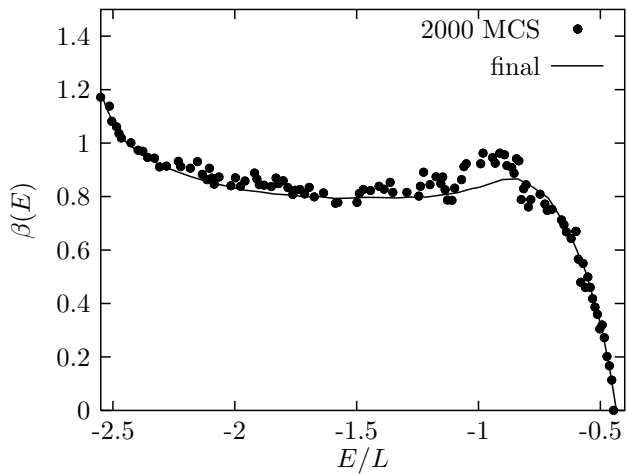}
	\caption{Symbols show the microcanonical inverse temperature $\beta(E)$ computed from the transition matrix accumulated over $2000$ MCS, for the six-state LR model ($\sigma=0.5$) containing $512$ spins. The estimate obtained from an interpolation scheme after the ultimate iteration is shown as a solid line for comparison.}
	\label{fig:beta_tmatrix}
\end{figure}
In view of the previously mentioned requirements on the estimation of $\beta(E)$, it is clear that, if one does not have a reliable guess of $\beta(E)$ at hand before the simulation starts, an efficient scheme must be devised in order to compute $\beta(E)$ in the early stage of the Wang-Landau algorithm. This is vital at this stage, because the exceedingly noisy estimate of the density of states makes it more likely to obtain under- or over-estimated values for $\beta(E)$.
An efficient approach in this regards relies on transition matrices \cite{Wang1999, Swendsen2004}. This method produces highly precise estimates of $\beta(E)$, although it has an inherently higher cost in term of computer load. The starting point is the Broad Histogram equation \cite{Wang2002,Oliveira1996}:
$$
n(E) T_{\infty}(E\zu E') =n(E') T_{\infty}(E'\zu E),
$$
where $T_\infty(E\zu E')$ is the transition matrix element between energy levels $E$ and $E'$ (also denoted as $\meanval{N(\sigma,E'-E)}_E$ in \cite{Oliveira1996}). This quantity contains the microcanonical average at energy $E$ of the number of potential single-spin moves from a state $\sigma$ of energy $E$ to a state $\sigma'$ of energy $E'$. It is estimated by accumulating a double-entry histogram $h(E,\Delta E)$ containing the number of potential moves from $E$ to $E+\Delta E$ each time the energy level $E$ is visited.
Long-range interactions lead to energy levels which are irregularly spaced, with in particular a few gaps in the vicinity of the ground state \cite{ReynalDiep2004a}, and it is necessary to choose an axis bin small enough to minimize discretization errors, and at the same time sufficiently large to contain at least a handful of entries.  In this case, $T_\infty(E\zu E')$ varies sufficiently smoothly for the following approximation scheme to be valid:
$$
\beta(E) =
\frac{1}{\Delta E} \ln \frac{T_{\infty}(E\zu E+\Delta E)}{T_{\infty}(E\zu E-\Delta E},
$$
where the actual estimate is obtained by weighted-averaging over several values of $\Delta E$. As illustrated in Fig.~\ref{fig:beta_tmatrix} for the six-state LR chain, the estimation of $\beta(E)$ from the transition matrix elements is reliable already after $2000$ MCS, which roughly corresponds to $50$ round-trips between the upper and the lower energy range.
For long-range models, each estimation of the number of potential moves requires of order $L^{2D}$ operations (as opposed to $L^D$ for nearest-neighbor interactions). However, we have shown in Sec.~\ref{sec:lr_optimization} that a single cluster update can demand as little as $O(L^D \ln L^D)$ operations when long-range specific optimizations are carried out; hence estimation schemes based on transition matrices partly scupper the benefits of these optimizations, and should therefore be employed only as a bootstrap procedure when nothing is known yet about the microcanonical temperature. Conversely, models with nearest-neighbor interactions do not undergo such a drawback, and make the transition matrix approach a perfectly transparent one from the viewpoint of algorithm complexity.

\subsection{Efficient predictors for the Wang-Landau algorithm}

\begin{figure}[h]
	\centering
	\InsertFig{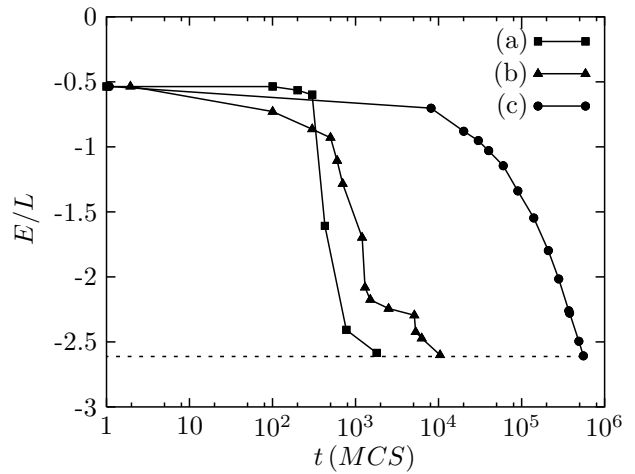}
	\caption{The graph shows the number of MCS needed to reach the ground-state  (dashed horizontal line) of the six-state Potts chain ($\sigma=0.5$ and $L=512$) for an initially unknown density of states,  using three distinct schemes: (a) and (b) predictor based on $\beta(E)$, local- and collective-update algorithms respectively; (c) no predictor ($S(E)=0,\  \forall E$), local-update algorithm.} 
	\label{fig:predictor_comparison_q6s5L9}
\end{figure}
Finally, we discuss how $\beta(E)$ can be used as an efficient predictor during the early stage of the Wang-Landau algorithm when nothing is known about the density of states. In the original implementation of this algorithm, we start with $S(E)=0$ for all energy levels, and simply increment $S(E)$ by the modification factor $\ln f$ each time the corresponding energy level is visited. One of the main drawbacks of this approach is that the Markov chain tends to wander around a fairly long time in the upper energy range, until eventually enough visits have been recorded in the histogram for the system to start exploring low-energy levels. This point has already been mentioned in \cite{Zhou2003}, where it was suggested that starting with a good initial guess of $S(E)$ was more efficient in terms of the number of histogram entries required to reach the final estimate, than performing a multi-range run with no initial guess at all. To circumvent this drawback when no initial guess is available, we therefore propose to use $\beta(E)$ to predict $S(E)$ for energy levels that are visited for the first time, and thus for which $S(E)$ is not available (i.e., it is set to $S(E)=0$ in the original implementation of the Wang-Landau algorithm). A linear prediction scheme turned out to sufficiently efficient for our purpose. As illustrated in Fig.~\ref{fig:predictor_comparison_q6s5L9}, using a predictor brings about a gain of three orders of magnitude in the time needed to reach the ground state. Our method and the single-spin update method lead similar performance, with however a slightly better behavior when cluster updates are used. We note that the Markov chain stays initially somewhat longer in the upper energy range when cluster updates are used, since a good estimate of $\beta(E)$ is needed to build the clusters with the correct bond probabilities. We think that this approach would prove particularly useful when the characteristics of the model makes it impossible to obtain an initial guess of $S(E)$ from simulations at smaller lattice sizes, e.g., in the presence of disorder or when the long-range interaction experience a slow decay.

\section{Conclusion}\label{sec:conclusion}

In conclusion, we have developed a new Monte Carlo method which combines in an  efficient and straightforward way the benefits of flat histogram algorithms with the fast-decorrelating capabilities of cluster updates. It is suited for spin models with any number of interaction between spins. Our formulation is versatile, and the method can be applied to a variety of density of states estimation schemes, including the Wang-Landau algorithm, Berg's recursion scheme or the transition matrix method.
We have shown that using the microcanonical temperature to compute cluster bond probabilities leads to a drastic reduction in effective autocorrelation times,  tunneling times and equilibration times. In the context of the Wang-Landau implementation, the reduced correlation between successive binning of the energy histogram yields a lower error in the estimation of the density of states, and as a result more reliable estimates of thermodynamic averages. 
Several schemes for the estimation of the microcanonical temperature were proposed, amongst which an efficient procedure which harnesses the power of the transition matrix method, and allows us to bootstrap the algorithm even if nothing is known initially about the density of states.
Finally, we carefully examined the precision of our method in the case of spin models with power-law decaying interactions. Here, our method proves all the more powerful that it is able to reduce the algorithm complexity to that of a short-range model having the same number of spins. This allowed us to study several finite-size effects at large lattice sizes, otherwise largely out of reach of conventional local-update implementations. In particular, we found out that the interface free energy scales perfectly well with a power of the lattice size, yet with a non-integer exponent which lies between $0$ and $1$. This, we think, is accounted for by the complex topology of the phases in coexistence in long-range models. A more detailed study, including a deeper insight into the topological properties of the generated clusters and the estimation of correlation lengths at large lattice sizes, would be very promising. We think that our method clearly draws this challenge within computation range.

\end{document}